\documentclass[11pt,a4paper]{article}
\usepackage[T1]{fontenc}
\usepackage[utf8]{inputenc}
\usepackage{lmodern}
\usepackage[margin=2.5cm]{geometry}
\usepackage{amsmath,amssymb,amsfonts}
\usepackage{graphicx}
\usepackage{bm}
\usepackage{xcolor}
\usepackage{array}
\usepackage{tabularx}
\usepackage{ragged2e}
\usepackage[section]{placeins}
\usepackage[final]{microtype}
\usepackage{hyperref}
\newtheorem{theorem}{Theorem}

\hypersetup{
  colorlinks=true,
  linkcolor=blue,
  citecolor=blue,
  urlcolor=blue
}
\begin{document}

\justifying

\title{Nodal algebraic curves and entropy diagnostics in degenerate two-dimensional harmonic-oscillator shells}

\author{C. A. Escobar, H. Olivares-Pilon, and A. M. Escobar-Ruiz\\[0.5em]
\small Departamento de F\'{i}sica, Universidad Aut\'onoma Metropolitana Unidad Iztapalapa,\\
\small San Rafael Atlixco 186, 09340 Ciudad de M\'exico, M\'exico\\[0.5em]
\small Corresponding author: \href{mailto:carlos.escobar@xanum.uam.mx}{carlos.escobar@xanum.uam.mx}}
\date{}

\maketitle

\begin{abstract}
\justifying
Degeneracy allows the nodal structure of a quantum state to vary without changing its energy. We study this effect for real superpositions in the fixed-energy shells of the two-dimensional isotropic harmonic oscillator. In Cartesian coordinates $(x,y)$, any such state in the \(N\)th shell has the parametrized form \(\psi_N(x,y;\mathbf{c})=\psi_0(x,y)\,P_N(x,y;\mathbf{c})\), where $\psi_0$ is the non-vanishing ground state and the real coefficients \(\mathbf{c}\) determine the polynomial factor \(P_N\). The nodal set is exactly the algebraic curve \(P_N=0\). Thus, varying \(\mathbf{c}\) reshapes this curve at fixed energy. We show that its topology can change only when it develops a finite singular point, \(P_N=\nabla P_N=0\), or when two asymptotic nodal directions merge.

We characterize this geometry using three diagnostics: the nodal-domain entropy \(S_{\rm dom}\), the Cartesian mutual information \(I(x;y)\), and the entropic uncertainty sum \(S_r+S_p\), which respectively probe probability redistribution among nodal domains, coordinate correlations, and global position--momentum delocalization. The lowest shells reveal a clear hierarchy. For \(N=1\), mixing only rotates a nodal line. Along a representative \(N=2\) path, a closed conic passes through parallel lines into a hyperbola-type curve; \(S_{\rm dom}\) clearly detects this transition, while \(S_r+S_p\) remains smooth. For \(N=3\), merging asymptotic directions brings smooth cubic branches close together and enhances both \(S_{\rm dom}\) and \(I(x;y)\). These signatures are accessible in Hermite–Gaussian structured light and nearly isotropic trapped oscillators. 
\end{abstract}


\section{Introduction}
\label{sec:intro}

In one dimensional quantum mechanics, the relation between energy ordering and nodal structure is
particularly rigid. Let us consider a regular Sturm--Liouville problem on an interval
\((a,b)\), eigenstates ordered by increasing eigenvalue and labeled by quantum number
\(n=1,2,\ldots\). The \(n\)th real normalized eigenfunction $\psi_n(x)$ has \(n-1\)
interior zeros, and these zeros divide the interval into exactly \(n\) connected
nodal domains~\cite{CourHil}. Denote these intervals by \(I_k^{(n)}\),
\(k=1,\ldots,n\). On each interval the wave function has a definite sign. Thus,
once the spectral index \(n\) is fixed, both the number and ordering of the
nodal intervals are fixed. In one dimension, there is no possibility for
nodal branches to reconnect or for closed nodal regions to appear.

The nodal partition also provides a natural description of how probability
is shared among the different regions. With
\begin{equation}
\rho_n(x)=|\psi_n(x)|^2 ,
\end{equation}
the probability contained in the \(k\)th nodal interval is given by
\begin{equation}
p^{(n)}_k
=
\int_{I^{(n)}_k}\rho_n(x)\,dx,
\qquad
\sum_{k=1}^{n}p^{(n)}_k=1 .
\label{eq:1d-domain-weights}
\end{equation}
Two complementary entropies can then be defined. The configuration-space
Shannon entropy,
\begin{equation}
S_r^{(n)}
=
-\int_a^b \rho_n(x)
\ln\!\left[\ell_0\rho_n(x)\right]dx ,
\label{eq:1d-shannon}
\end{equation}
where $\ell_0$ fixes the reference length entering
the differential entropy, measures the overall spatial spread of the probability density. By contrast,
the nodal-domain entropy,
\begin{equation}
S^{(n)}_{\rm dom}
=
-\sum_{k=1}^{n}p^{(n)}_k\ln p^{(n)}_k ,
\label{eq:1d-domain-entropy}
\end{equation}
measures only how the total probability is distributed among the nodal
domains. It satisfies
\[
0\leq S^{(n)}_{\rm dom}\leq \ln n,
\]
with the upper bound reached when all \(n\) domains carry equal probability.

The infinite square well provides a useful and transparent benchmark. For a well of width
$L$ centered at the origin, the $n$th eigenstate contains $n$ identical
half-wave lobes separated by $n-1$ interior nodes. Each nodal domain
carries probability $1/n$, giving
\begin{equation}
S_{\rm dom}^{(1)}=0,\qquad
S_{\rm dom}^{(2)}=\ln 2,\qquad
S_{\rm dom}^{(3)}=\ln 3,
\end{equation}
for the three lowest states, and $S_{\rm dom}^{(n)}=\ln n$ in general.
By contrast, direct integration yields
\begin{equation}
S_r^{(n)}=\ln\left(\frac{2L}{\ell_0}\right)-1
\end{equation}
for every eigenstate. Excitation partitions the probability into an
increasing number of sign-definite lobes without enlarging the accessible
interval. Thus, in this example, $S_{\rm dom}$ resolves the nodal
partition, whereas $S_r$ is independent of $n$. 

What changes in higher dimensions is the direct connection between spectral
ordering and nodal geometry. In two dimensions, nodal sets are generally
curves. They may form closed loops, extend to infinity, intersect, or change
their connectivity. General nodal theorems provide constraints, such as
Courant bounds, but do not uniquely determine the number or topology of the
nodal domains. The freedom is even greater inside a degenerate eigenspace:
the coefficients of a superposition may vary while the energy remains
unchanged, allowing the nodal curve to deform at fixed energy
~\cite{CourHil,CharronL2025,Karp_1989}. The problem is no longer
of counting zeros, but of determining when nodal curves can reconnect,
split, or change their asymptotic structure. This brings the problem into
contact with real algebraic curves and singularity theory
~\cite{VladimirIArnold_1989,VladimirIArnold_2011}.

We study this phenomenon in the two-dimensional isotropic harmonic oscillator $V(x,y)=\frac{1}{2}m\,\omega^2\,(x^2+y^2)$. Its spectrum is organized into shells labeled by the principal quantum number $N=n_x+n_y=0,1,2,\ldots$. The $N$th shell has energy $E_N=\hbar\,\omega\,(N+1)$ and contains $N+1$ degenerate Cartesian product states. A normalized real state in this shell is specified by a coefficient vector $\mathbf{c}=(c_0,\ldots,c_N)$. Because all basis states share the same Gaussian envelope (the ground state eigenfunction), the remaining spatial dependence is described by a real polynomial $P_N(x,y;\mathbf{c})$ of degree $N$. The coefficients therefore deform the polynomial, and hence its nodal geometry, without changing the energy. The formal construction and normalization conventions are given in Sec.~\ref{sec2}. 

The Gaussian envelope has no zeros, so the nodal structure is determined entirely by the polynomial factor $P_N(x,y;\mathbf{c})$. As discussed in Sec.~\ref{sec:cart_fixed_shell}, its topology can change through two distinct mechanisms. The first is a finite singularity, where nodal branches meet, cross, or become tangent. The second is an asymptotic degeneracy, where two nodal directions merge at infinity. Away from these exceptional configurations, the curve deforms smoothly without changing its topology. Finite and asymptotic degeneracies identify the parameter values at which nodal restructuring can occur within a fixed energy shell~\cite{FultonAlgebraicCurves,FischerPlaneCurves,MilnorSingularPoints}.

These geometric criteria locate possible nodal transitions, but they do not describe how probability is redistributed when the curve changes. Information-theoretic quantities provide this complementary description. Differential Shannon entropies and mutual information have been used to resolve coordinate correlations, compare position- and momentum-space behavior, and reveal the influence of nodes on two-dimensional probability densities~\cite{SchuergerEngel2023Nodes,SchuergerEngel2023Entropy}. This leads to the central physical question addressed here: how does nodal restructuring within a degenerate fixed-energy shell appear in entropy and correlation measures?

The nodal structure of harmonic-oscillator eigenstates and their superpositions has been investigated from geometric, spectral, and semiclassical perspectives~\cite{BH2015,10.1093/imrn/rny290}, while the associated entropic diagnostics have largely been developed independently~\cite{PhysRevA.50.3065,DJTI2020}. Here we bring these two directions together. The algebraic criteria determine where a nodal curve can reorganize, whereas the information measures quantify the accompanying redistribution of probability and Cartesian correlations. The energy selects the oscillator shell; the superposition coefficients determine its internal nodal geometry.

A crucial restriction is that a fixed oscillator shell does not generate an
arbitrary real algebraic curve of degree \(N\). Instead, the Hermite basis
produces a constrained hierarchy: lines for \(N=1\), centered conics for
\(N=2\), constrained odd cubics for \(N=3\), and constrained even quartics
for \(N=4\). The explicit low-shell families, including the projective
discriminant of the cubic shell, are collected in the Supplemental Material.
Within each family, finite singularities and repeated asymptotic directions
identify the parameter values at which the nodal geometry may reorganize.

We use three complementary diagnostic roles. The nodal-domain entropy
\(S_{\rm dom}\) measures how probability is shared among connected nodal
regions. The Cartesian mutual information \(I(x;y)\) measures correlations
between the two coordinates. Finally, \(S_r\) and the entropic uncertainty
sum \(S_r+S_p\) measure spatial and position--momentum delocalization,
respectively~\cite{Wehner_2010,VMajerník_1996}. The lowest nontrivial shells
show progressively richer behavior. For \(N=1\), changing the superposition
only rotates a single nodal line. For \(N=2\), a representative path takes
the nodal curve from a closed conic, through parallel lines, to a
hyperbola-type curve. This transition is clearly detected by
\(S_{\rm dom}\), while \(S_r+S_p\) remains smooth. For \(N=3\), merging
asymptotic directions organizes regimes in which smooth cubic branches
approach one another closely, with enhanced responses in \(S_{\rm dom}\)
and \(I(x;y)\).

The same quantities can be reconstructed experimentally. In real-phase
Hermite--Gaussian structured light, the nodal curves are the dark lines of
the transverse intensity pattern, while near-field and far-field
measurements provide the position- and momentum-space distributions.
Analogous information can be obtained in nearly isotropic trapped motional
systems. These platforms provide direct settings in which nodal
restructuring and its information-theoretic signatures can be tested. The shell hierarchy and the corresponding diagnostic responses are
summarized in Table~\ref{tab:low-shell-hierarchy}.

\begin{table*}[t]
\caption{
Nodal geometry and information diagnostics in fixed oscillator shells.
Finite singularities and repeated asymptotic directions identify where the
nodal topology may change. The three diagnostics probe probability
redistribution among nodal domains, Cartesian correlations, and global
delocalization.
}
\label{tab:low-shell-hierarchy}

\centering
\small
\setlength{\tabcolsep}{4pt}
\renewcommand{\arraystretch}{1.15}

\begin{tabular}{
@{}c
>{\raggedright\arraybackslash}p{0.16\textwidth}
>{\raggedright\arraybackslash}p{0.24\textwidth}
>{\raggedright\arraybackslash}p{0.39\textwidth}
@{}
}
\hline\hline
Shell &
Nodal geometry &
Conditions for restructuring &
Diagnostic signature
\\
\hline

$N=1$
&
Single line, $P_1=0$
&
No finite singularity for a nonzero normalized state; the asymptotic
structure is unchanged.
&
The nodal line undergoes only a rigid rotation. The two nodal domains have
equal weights, so $S_{\rm dom}=\ln 2$ and $S_r+S_p$ are independent of the
superposition coefficients. $I(x;y)$ measures the orientation relative to
the Cartesian axes.
\\[0.5em]

$N=2$
&
Centered conic, $P_2=0$
&
Finite branch crossing or loss of rank of the quadratic part, producing the
parallel-line configuration.
&
Along the symmetric path, the nodal curve passes from a closed conic,
through parallel lines, to a hyperbola-type curve. $S_{\rm dom}$ detects the
change in the nodal partition, whereas $S_r+S_p$ remains smooth.
\\[0.5em]

$N=3$
&
Hermite-constrained cubic, $P_3=0$
&
Finite singularities of the full cubic or repeated asymptotic directions of
its leading homogeneous part.
&
Looped and close-branch geometries occur. Along the path studied, the close
approach of smooth branches is organized by the projective degeneracy rather
than by an interior finite singularity. $S_{\rm dom}$ and $I(x;y)$ show the
largest variations.
\\[0.5em]

General $N$
&
Hermite-constrained algebraic curve of degree at most $N$
&
Finite condition $P_N=\nabla P_N=0$ or repeated directions of the leading
homogeneous part $P_{N,\mathrm{top}}$.
&
Away from these conditions, the nodal set is topologically stable and the
nodal weights vary smoothly. $S_{\rm dom}$, $I(x;y)$, and $S_r+S_p$ probe
complementary aspects of the restructuring.
\\

\hline\hline
\end{tabular}
\end{table*}

\section{Two-Dimensional Isotropic Harmonic Oscillator}
\label{sec2}

For the two-dimensional isotropic harmonic oscillator, the fixed shell
\begin{equation}
N=n_x+n_y \ ,
\label{eq:N_def}
\end{equation}
has energy
\begin{equation}
E_N=\hbar\,\omega(N+1) \ ,
\label{eq:EN_def}
\end{equation}
and dimension $N+1$. A convenient basis is the Cartesian product basis
\begin{equation}
\begin{aligned}
\Phi_{n,N-n}(x,y)&=\phi_n(x)\,\phi_{N-n}(y)\quad ;\qquad
n=0,1,\ldots,N.
\end{aligned}
\label{eq:cart_basis}
\end{equation}
where $\phi_n$ are the one-dimensional oscillator eigenfunctions. Any real state in this shell can be written as
\begin{equation}
\psi_N(x,y)=\sum_{n=0}^{N} c_n\,\phi_n(x)\phi_{N-n}(y),
\qquad
\sum_{n=0}^N c_n^2=1,
\label{eq:psi_shell_expansion}
\end{equation}
with real coefficients $c_n$ defined up to an overall sign. This restriction to real coefficients is essential in the present
work: it ensures that the nodal set is described by the single real
algebraic equation $P_N(x,y)=0$. Generic complex superpositions
would require the simultaneous vanishing of the real and imaginary
parts of the wavefunction and may contain phase singularities. Using the Hermite form
\begin{equation}
\phi_n(x)=\mathcal{N}_n\,H_n(\sqrt{\alpha}\,x)\,e^{-\alpha x^2/2},
\qquad
\alpha=\frac{m\,\omega}{\hbar},
\label{eq:phi_hermite}
\end{equation}
one obtains

\begin{equation}
\psi_N(x,y)=\mathcal C_N\,e^{-\alpha\,r^2/2}\,P_N(x,y),
\qquad r^2=x^2+y^2,
\label{eq:gauss_poly}
\end{equation}
where \(\mathcal C_N\neq0\) is independent of \(x,y\) and $P_N(x,y)$ is a real
polynomial of total degree $N$, defined up to an overall nonzero constant. The
factor \(\mathcal C_N\) can always be absorbed into \(P_N\); below we keep it
separate only when convenient for normalization. Hence, the shell dependence
enters through the polynomial factor $P_N$, while the Gaussian envelope is fixed.

\paragraph*{Hidden $SU(2)$ structure.}
The isotropic oscillator carries a dynamical $SU(2)$ symmetry within each degenerate shell. Writing
\[
H=\hbar\,\omega\,(a_x^\dagger a_x+a_y^\dagger a_y+1) \ ,
\]
and introducing the Schwinger generators
\[
J_+=a_x^\dagger a_y,\qquad
J_-=a_y^\dagger a_x,\qquad
J_z=\tfrac12(a_x^\dagger a_x-a_y^\dagger a_y),
\]
with $J_x=(J_++J_-)/2$ and $J_y=(J_+-J_-)/(2i)$, one has $[H,J_i]=0$, and the operators $J_i$ satisfy the $su(2)$ commutation relations. The shell $N$ realizes the spin-$j=N/2$ irreducible representation of $SU(2)$, with the product basis $\{\Phi_{n_x,n_y}\}$ identified with $\{|j,m\rangle\}$ through
\[
m=\frac{n_x-n_y}{2}.
\]

From this viewpoint, varying the coefficients $\{c_n\}$ amounts to moving within an $SU(2)$ multiplet, which provides a symmetry interpretation of the one-parameter families considered below.

\section{Fixed-shell geometry and information measures}
\label{sec:cart_fixed_shell}

\subsection{Nodal set and degeneracy criteria}
\label{subsec:cart_nodal}

Since the Gaussian envelope in Eq.~\eqref{eq:gauss_poly} is strictly positive, the zeros of $\psi_N$ coincide with those of the polynomial factor:
\begin{equation}
P_N(x,y)=0.
\label{eq:nodal_set}
\end{equation}
For generic coefficients $\{c_n\}$, this defines a real algebraic curve of degree at most $N$. Qualitative changes in the nodal geometry can occur when the curve develops a
finite singular point,
\begin{equation}
P_N(x_c,y_c)=0,
\qquad
\nabla P_N(x_c,y_c)=\mathbf{0},
\label{eq:bifurcation_condition}
\end{equation}
or when the asymptotic nodal directions of the highest-degree homogeneous
part become non-simple.

\subsection{Configuration-space information measures}
\label{subsec:cart_info}

In two dimensions, the normalized spatial density is
\begin{equation}
\rho(x,y)=
\frac{|\psi_N(x,y)|^2}
{\displaystyle\iint_{\mathbb{R}^2}
|\psi_N(x,y)|^2\,dx\,dy}.
\label{eq:rho_def}
\end{equation}
The geometric criteria identify where the nodal pattern may reorganize. To determine how this reorganization affects the probability distribution, we introduce the information-theoretic measures used below.

Throughout the two-dimensional analysis, a single fixed convention is
used for the differential entropies. For numerical calculations, we
adopt the natural oscillator scales
\begin{equation}
\ell_{\mathrm{osc}}=\sqrt{\frac{\hbar}{m\omega}},
\qquad
p_{\mathrm{osc}}=\frac{\hbar}{\ell_{\mathrm{osc}}}
=\sqrt{m\hbar\omega}. \nonumber
\label{eq:osc_scales}
\end{equation}
Equivalently, one may use the dimensionless variables
\[
X=\frac{x}{\ell_{\mathrm{osc}}}=\sqrt{\alpha}\,x,
\qquad
Y=\frac{y}{\ell_{\mathrm{osc}}}=\sqrt{\alpha}\,y,
\qquad
\Pi_x=\frac{p_x}{p_{\mathrm{osc}}},
\qquad
\Pi_y=\frac{p_y}{p_{\mathrm{osc}}}.
\]

In these dimensionless variables, the numerical probability densities
and the arguments of the logarithms are dimensionless. When dimensional
variables are retained in the analytic expressions below, the
differential entropies carry the usual additive dependence on the
chosen length and momentum units. The compact notation
$\ln\rho$, $\ln\rho_x$, $\ln\rho_y$, and $\ln\widetilde{\rho}$ is
therefore understood within this fixed unit convention. Since the same
convention is used along each coefficient path, these additive terms
are independent of the superposition coefficients and do not affect
the coefficient-dependent variations or the Cartesian mutual
information. Accordingly, all numerical calculations are performed in
these dimensionless oscillator variables, with $\alpha=1$.

\begin{enumerate}
\item Configuration-space Shannon entropy $S_r$
 \begin{equation}
S_r=-\iint_{\mathbb{R}^2}\rho(x,y)\ln\rho(x,y)\,dx\,dy.
\label{eq:entropy_2d}
\end{equation}
\item The Cartesian mutual information given by 
\begin{equation}
I(x;y)=S_x+S_y-S_r\ge 0,
\label{eq:mutual_information}
\end{equation}
is defined in terms of $S_x$ and $S_y$,
\begin{equation}
\begin{aligned}
S_x&=-\int\rho_x\ln\rho_x\,dx\quad ,\qquad
S_y&=-\int\rho_y\ln\rho_y\,dy,
\end{aligned}
\end{equation}
where the marginal distributions are 
\begin{equation}
\rho_x(x)=\int_{-\infty}^{\infty}\rho(x,y)\,dy\quad , \qquad
\rho_y(y)=\int_{-\infty}^{\infty}\rho(x,y)\,dx,
\label{eq:marginals}
\end{equation}
respectively. For a single particle in two dimensions, \(I(x;y)\) is not an
entanglement measure between two particles. Rather, it quantifies
statistical correlations between the two Cartesian degrees of freedom
in the position representation. In the oscillator basis, these correlations are induced by coherent superpositions inside the degenerate shell. Therefore, \(I(x;y)\) provides a basis-dependent, but physically transparent, measure of coordinate nonseparability within the fixed Cartesian oscillator representation.
\item Nodal-domain entropy $S_{\mathrm{dom}}$.
Let $\{\Omega_k\}$ be the connected components of
\begin{equation}
\mathbb{R}^2\setminus\{(x,y):\psi_N(x,y)=0\},
\label{eq:nodal_domains}
\end{equation}
that is, the maximal connected open sets on which $\psi_N$ has definite sign. Their probability weights are
\begin{equation}
p_k=\int_{\Omega_k}\rho(x,y)\,dx\,dy,
\qquad
\sum_k p_k=1,
\label{eq:pk_def}
\end{equation}
and the associated entropy is
\begin{equation}
S_{\mathrm{dom}} =-\sum_k p_k\ln p_k.
\label{eq:Sdom_def}
\end{equation}
Unlike $S_r$, this quantity is tied directly to the nodal partition, since a change in nodal topology reorganizes the set $\{p_k\}$ itself. The integrals in Eq.~\eqref{eq:pk_def} converge for all fixed-shell states because $\rho$ has Gaussian decay.
\item Entropic uncertainty sum $S_r+S_p$.
Given the momentum-space wavefunction $\tilde{\psi}_N(p_x,p_y)$, the normalized momentum density
$\tilde{\rho}(p_x,p_y)$ is
\begin{equation}
\tilde{\rho}(p_x,p_y)=\frac{|\tilde{\psi}_N(p_x,p_y)|^2}
{\iint_{\mathbb{R}^2}|\tilde{\psi}_N(p_x,p_y)|^2\,dp_x\,dp_y},
\label{eq:rho_p_def}
\end{equation}
and the corresponding momentum-space Shannon entropy takes the form
\begin{equation}
S_p
=
-\iint_{\mathbb{R}^2}\tilde{\rho}(p_x,p_y)\ln\tilde{\rho}(p_x,p_y)\,dp_x\,dp_y.
\label{eq:entropy_p}
\end{equation}
The entropic uncertainty sum
\begin{equation}
S_r+S_p,
\label{eq:entropic_sum}
\end{equation}
measures overall delocalization across configuration and momentum space.
\end{enumerate}

\section{Regular evolution of nodal patterns}

Changing the superposition coefficients within a fixed oscillator shell
deforms the wavefunction without changing its energy. Two regimes are
then possible. If no nodal branches meet at a finite point and no
asymptotic directions merge, the nodal pattern evolves smoothly: no
domain is created or destroyed, and the associated probabilities vary
regularly. A change of nodal topology requires one of these conditions
to fail. We now state this result for an arbitrary shell.

Let \(J\subset\mathbb R\) be a compact interval and consider the
\(C^2\) family of normalized real states
\begin{equation}
\begin{aligned}
\psi_t(x,y)
&=
\sum_{n=0}^{N}c_n(t)\,\Phi_{n,N-n}(x,y),
\qquad
c_n\in C^2(J;\mathbb R),
\qquad
\sum_{n=0}^{N}c_n(t)^2=1 ,
\end{aligned}
\label{eq:regular-family-shell}
\end{equation}
where $\Phi_{n,N-n}(x,y)$ is given by~(\ref{eq:cart_basis}).
Using the Gaussian--polynomial representation,
\begin{equation}
\psi_t(x,y)
=
e^{-\alpha(x^2+y^2)/2}P_t(x,y),
\label{eq:regular-family-gaussian}
\end{equation}
see Eq.~\eqref{eq:gauss_poly}, we define the nodal curve and probability
density by
\begin{equation}
Z_t
=
\{(x,y)\in\mathbb R^2:P_t(x,y)=0\},
\qquad
\rho_t=|\psi_t|^2 .
\label{eq:regular-family-Zt-rhot}
\end{equation}
The normalization in Eq.~\eqref{eq:regular-family-shell} and the
orthonormality of the shell basis imply
\(\int_{\mathbb R^2}\rho_t\,dx\,dy=1\).

Let \(P_t^{(N)}\) denote the homogeneous degree-\(N\) part of \(P_t\),
which controls the nodal directions at large distance. We restrict the
discussion to intervals on which \(P_t^{(N)}\) is not identically zero;
a loss of degree is itself a degeneracy. For every \(t\in J\), we assume:

\begin{enumerate}
\item[(i)] the equations
\begin{equation}
P_t(x,y)=0,
\qquad
\nabla P_t(x,y)=0
\label{eq:regular-assumption-sing}
\end{equation}
have no common real solution;

\item[(ii)] the angular function
\begin{equation}
f_t(\theta)
:=
P_t^{(N)}(\cos\theta,\sin\theta)
\label{eq:regular-assumption-ft}
\end{equation}
has only simple real zeros.
\end{enumerate}

Condition (i) excludes finite crossings and tangencies of nodal branches.
Condition (ii) prevents two asymptotic nodal directions from merging.
Together, they exclude the two mechanisms through which the nodal
topology can change.

\begin{theorem}
\label{thm:regular-family}
Under the fixed-degree assumption and conditions {\rm (i)}--{\rm (ii)},
the following statements hold.

\begin{enumerate}
\item[(a)] For every \(t\in J\), \(Z_t\) is a smooth, properly embedded
one-dimensional submanifold of \(\mathbb R^2\). For each \(t_0\in J\),
there is a neighborhood \(U\subset J\) and a
\(C^1\) family of \(C^1\) diffeomorphisms
\begin{equation}
H_t:\mathbb R^2\to\mathbb R^2,
\qquad
t\in U,
\qquad
H_{t_0}=\mathrm{id},
\label{eq:regular-ambient-isotopy}
\end{equation}
such that
\begin{equation}
H_t(Z_{t_0})=Z_t .
\label{eq:regular-isotopy-relation}
\end{equation}
Thus, after relabeling the connected components if necessary,
\begin{equation}
\mathbb R^2\setminus Z_t
=
\bigsqcup_{k=1}^{M}\Omega_k(t),
\label{eq:regular-nodal-domains}
\end{equation}
where the number \(M\) of nodal domains is constant on \(J\).

\item[(b)] The probability carried by each nodal domain,
\begin{equation}
p_k(t)
=
\int_{\Omega_k(t)}
\rho_t(x,y)\,dx\,dy,
\qquad
k=1,\ldots,M,
\label{eq:regular-pk}
\end{equation}
is a positive \(C^1\) function of \(t\), with
\begin{equation}
\sum_{k=1}^{M}p_k(t)=1.
\label{eq:regular-pk-sum}
\end{equation}

\item[(c)] The nodal-domain entropy
\begin{equation}
S_{\rm dom}(t)
=
-\sum_{k=1}^{M}p_k(t)\ln p_k(t)
\label{eq:regular-Sdom}
\end{equation}
is of class \(C^1\) on \(J\).

\item[(d)] The configuration-space Shannon entropy
\begin{equation}
S_r(t)
=
-\iint_{\mathbb R^2}
\rho_t(x,y)\ln\rho_t(x,y)\,dx\,dy
\label{eq:regular-Sr}
\end{equation}
is continuous on \(J\). With the Fourier convention and dimensional
variables used here, all states in a fixed shell acquire the same
Fourier phase, and the momentum density is
\[
\widetilde\rho_t(p_x,p_y)
=
\frac{1}{(m\omega)^2}
\rho_t\!\left(
\frac{p_x}{m\omega},
\frac{p_y}{m\omega}
\right).
\]
It follows that
\[
S_p(t)=S_r(t)+2\ln(m\omega).
\]
In dimensionless oscillator variables, \(S_p(t)=S_r(t)\). Therefore,
the entropic sum \(S_r(t)+S_p(t)\) is continuous on \(J\).
\end{enumerate}
\end{theorem}

The physical content of Theorem~\ref{thm:regular-family} is direct:
away from finite branch contacts and mergers of asymptotic directions,
the nodal pattern can bend and move but cannot reconnect. Its number of
domains remains fixed, their probabilities change smoothly, and
\(S_{\rm dom}\) has no singular behavior. Consequently, every
topology-changing event discussed below must be associated with a
finite singularity, an asymptotic degeneracy, or a loss of polynomial
degree. The proof is given in Appendix~\ref{proof}.

\section{The $N=1$ shell: linear nodal geometry and coefficient-independent domain entropy}
\label{sec:N1_structural}

\subsection{Wavefunction and fixed-shell polynomial}

The $N=1$ shell is spanned by $\Phi_{10}=\phi_1(x)\phi_0(y)$ and
$\Phi_{01}=\phi_0(x)\phi_1(y)$. A general real state is
\begin{equation}
\psi_1(x,y)=a\,\Phi_{10}+b\,\Phi_{01},
\qquad
a^2+b^2=1.
\label{eq:N1_general}
\end{equation}
Using
\begin{equation}
\label{phi01}
\phi_0(x)=\left(\frac{\alpha}{\pi}\right)^{1/4}e^{-\alpha x^2/2},
\qquad
\phi_1(x)=\left(\frac{\alpha}{\pi}\right)^{1/4}\sqrt{2\alpha}\,x\,e^{-\alpha x^2/2},
\end{equation}
one finds the Gaussian--polynomial form $\psi_1=e^{-\alpha r^2/2}P_1(x,y)$, where
\begin{equation}
P_1(x,y)=\left(\frac{\alpha}{\pi}\right)^{1/2}\sqrt{2\alpha}\,(ax+by),
\end{equation}
so the polynomial factor is linear.

\subsection{Nodal set and absence of bifurcation}

Since the Gaussian envelope is strictly positive, the nodal set is determined
by
\begin{equation}
    ax+by=0 .
\end{equation}
Thus, every nontrivial \(N=1\) state has a single straight nodal line through
the origin. Up to an irrelevant nonzero factor, \(P_1\propto ax+by\), so
\(\nabla P_1\propto(a,b)\). Since \(a^2+b^2=1\), the gradient never vanishes,
and the conditions (\ref{eq:bifurcation_condition}) with $N=1$
cannot be satisfied. Hence, the \(N=1\) shell contains no nodal bifurcation;
coefficient variation only rotates the line.

\subsection{Exact probability density and normalization}

The normalized density is
\begin{equation}
\rho_1(x,y)=|\psi_1(x,y)|^2
=
\frac{2\alpha^2}{\pi}(ax+by)^2 \,e^{-\alpha\,r^2}.
\label{eq:N1_rho_explicit}
\end{equation}
Introducing rotated coordinates
\begin{equation}
u=ax+by,
\qquad
v=-bx+ay,
\label{eq:N1_rot}
\end{equation}
gives
\begin{equation}
\rho_1(u,v)=\frac{2\alpha^2}{\pi}u^2 e^{-\alpha(u^2+v^2)},
\end{equation}
showing that all dependence on $(a,b)$ amounts to a rigid rotation of a single
universal profile.

\subsection{Shannon entropy and entropic sum}

The configuration-space Shannon entropy $S_r$, defined in (\ref{eq:entropy_2d}),
is rotationally invariant and independent of $(a,b)$. Direct evaluation
gives
\begin{equation}
S_r=\ln\!\left(\frac{2\pi}{\alpha}\right)+\gamma,
\label{eq:N1_entropy_final}
\end{equation}
where $\gamma$ is Euler's constant.

In momentum space, the Fourier transform of
Eq.~\eqref{eq:N1_general} has the same fixed-shell structure. Writing
$p_0=\sqrt{m\hbar\omega}$, the normalized momentum density is
\begin{equation}
\rho_p(p_x,p_y)=
\frac{2}{\pi p_0^4}(ap_x+bp_y)^2
e^{-(p_x^2+p_y^2)/p_0^2},
\label{eq:rho_p_N1_short}
\end{equation}
which is again reduced by the rotation \eqref{eq:N1_rot} to a
coefficient-independent form. The entropy (\ref{eq:entropy_p}) reads
\begin{equation}
S_p
=\ln(2\pi p_0^2)+\gamma.
\label{eq:Sp_N1_short}
\end{equation}
Therefore,
\begin{equation}
S_r+S_p
=
2\gamma+2\ln(2\pi\hbar),
\label{eq:N1_sum_short}
\end{equation}
or, in the dimensionless convention $\hbar=1$,
\begin{equation}
S_r+S_p=2\gamma+2\ln(2\pi).
\label{eq:N1_sum_dimless_short}
\end{equation}
Along the path $a=\sqrt{1-t^2}$, $b=t$, this sum is constant: varying the
coefficients only rotates the state.

\subsection{Nodal-domain entropy}

The line $ax+by=0$ divides the plane into two nodal domains. In the coordinates
\eqref{eq:N1_rot}, these are $u>0$ and $u<0$. Because
$\rho_1(u,v)\propto u^2e^{-\alpha(u^2+v^2)}$ is even in $u$, both domains have
weight $1/2$:
\begin{equation}
p_+=\int_{u>0}\rho_1\,du\,dv=\frac{1}{2},
\qquad
p_-=\int_{u<0}\rho_1\,du\,dv=\frac{1}{2}.
\end{equation}
Thus, direct evaluation of (\ref{eq:Sdom_def}) gives
\begin{equation}
S_{\mathrm{dom},\,N=1}
=
\ln 2,
\label{eq:N1_Sdom}
\end{equation}
independent of $(a,b)$.

\subsection{Mutual information and axis dependence}

The mutual information depends on the choice of coordinates. In the rotated
frame \((u,v)\), the density factorizes and $I(u;v)=0$; in the fixed Cartesian
frame \((x,y)\), one generally has $I(x;y)\neq 0$ unless the nodal line is
axis-aligned.

For later numerical use, the marginals (\ref{eq:marginals}) are
\begin{align}
\rho_{1,x}(x)
&=
\sqrt{\frac{\alpha}{\pi}}\,e^{-\alpha x^2}\bigl(2\alpha a^2x^2+b^2\bigr),
\qquad 
\rho_{1,y}(y)
=
\sqrt{\frac{\alpha}{\pi}}\,e^{-\alpha y^2}\bigl(2\alpha b^2y^2+a^2\bigr).
\label{eq:N1_marginal_y}
\end{align}

\begin{figure}[h]
\centering
\includegraphics[width=0.5\columnwidth]{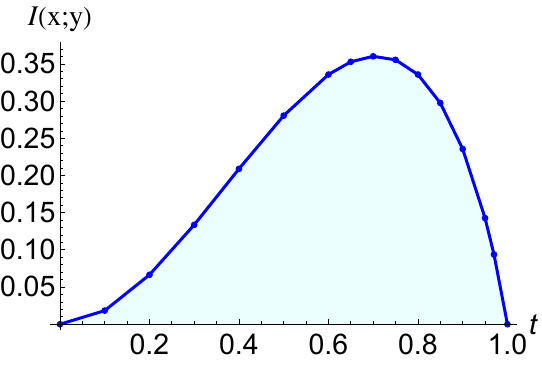}
\caption{Mutual information $I(x;y)$ versus the mixing parameter $t$ in the $N=1$ shell
($a=\sqrt{1-t^2}$, $b=t$). The peak near $t=1/\sqrt{2}$ corresponds to the configuration
for which the nodal line is maximally oblique relative to the fixed Cartesian axes.}
\label{fig:IxyN1}
\end{figure}

Writing $a=\sqrt{1-t^2}$ and $b=t$ makes the geometry explicit:
\begin{equation}
    ax+by=0
\quad\Longleftrightarrow\quad
\sqrt{1-t^2}\,x+t\,y=0.
\label{lnN1}
\end{equation}

Varying $t\in[0,1]$ simply rotates the nodal line~(\ref{lnN1}), and with it the density
$\rho_1$. Accordingly, $I(x;y)$ defined in (\ref{eq:mutual_information}) is a smooth function of $t$: it vanishes at
$t=0$ and $t=1$, when the nodal line is axis-aligned, and reaches its maximum
near $t=1/\sqrt{2}$, where the pattern is most oblique to the Cartesian axes
(see Fig.~\ref{fig:IxyN1}).

The $N=1$ shell therefore serves as a baseline case. Its nodal set is always a
single line, the two nodal domains always carry equal probability, and nonzero
$I(x;y)$ reflects only the choice of a non-adapted frame. Genuine nodal
restructuring first appears for $N\ge 2$.


\section{The \(N=2\) shell: conic nodal sets and the first topology-changing transition}
\label{sec:N2_structural}

\subsection{Wavefunction and quadratic polynomial}
\label{subsec:N2_wavefunction_poly}

The $N=2$ shell is spanned by $\Phi_{20}$, $\Phi_{11}$, and $\Phi_{02}$, so a
general real state $\psi_2(x,y)$ is
\begin{equation}
\psi_2(x,y)=a\,\Phi_{20}(x,y)+b\,\Phi_{11}(x,y)+c\,\Phi_{02}(x,y),
\label{eq:N2_superposition}
\end{equation}
with
\begin{displaymath}
    a^2+b^2+c^2=1.
\end{displaymath}

Using (\ref{eq:cart_basis}), with $\phi_0(x)$ and $\phi_1(x)$ from~(\ref{phi01}) and 
\begin{equation}
\phi_2(x)=\left(\frac{\alpha}{\pi}\right)^{1/4}\frac{1}{\sqrt{2}}\,
(2\alpha x^2-1)\,e^{-\alpha x^2/2},
\label{eq:N2_phi12}
\end{equation}
the normalized state can be written as
\[
\psi_2(x,y)=\sqrt{\frac{\alpha}{\pi}}\,
e^{-\alpha r^2/2}\,P_2(x,y),
\]
where
\begin{equation}
P_2(x,y)=
\sqrt{2}\,\alpha\,a\,x^2
+2\alpha\,b\,xy
+\sqrt{2}\,\alpha\,c\,y^2
-\frac{a+c}{\sqrt{2}}.
\label{eq:N2_poly_explicit}
\end{equation}

The prefactor $\sqrt{\alpha/\pi}$ is independent of $x$ and $y$ and has no effect on the nodal geometry. This is the normalization convention for
$P_2$ used below, in particular in the density of
Eq.~\eqref{eq:N2_rho_explicit}.

\subsection{The \(N=2\) transition: from a closed curve to open branches}
\label{subsec:N2_discriminant}

It is useful to write the centered conic as
\begin{equation}
P_2(x,y)=
\begin{pmatrix}x & y\end{pmatrix}
Q
\begin{pmatrix}x\\y\end{pmatrix}
+D,
\qquad
Q=
\begin{pmatrix}
A & B/2\\
B/2 & C
\end{pmatrix},
\label{eq:N2_Q_form}
\end{equation}
where
\begin{equation}
\begin{aligned}
A&=\sqrt{2}\,\alpha\,a,
\qquad
B=2\alpha\,b,
\qquad
C=\sqrt{2}\,\alpha\,c,
\qquad
D=-\frac{a+c}{\sqrt{2}}
=-\frac{A+C}{2\alpha}.
\end{aligned}
\label{eq:N2_Q_params}
\end{equation}
This form makes the two possible changes of nodal geometry transparent:
branches can cross at a finite point, or the conic can pass through a
pair of parallel lines as it changes from closed to open.

A finite crossing requires \(P_2=0\) and \(\nabla P_2=0\), as in
Eq.~\eqref{eq:bifurcation_condition}. Since
\begin{equation}
\nabla P_2(\mathbf{x})=2Q\mathbf{x},
\qquad
\mathbf{x}=
\begin{pmatrix}
x\\y
\end{pmatrix},
\end{equation}
a nonsingular \(Q\) has only one critical point, \(\mathbf{x}=0\). This
point lies on the nodal curve precisely when
\begin{equation}
D=0
\qquad\Longleftrightarrow\qquad
a+c=0 .
\label{eq:N2_affine_singularity}
\end{equation}
The nodal set then consists of two lines crossing at the origin.

A different transition occurs when the quadratic form loses rank:
\begin{equation}
\det Q
=
AC-\frac{B^2}{4}
=
\alpha^2(2ac-b^2).
\end{equation}
The corresponding condition is
\begin{equation}
b^2=2ac .
\label{eq:N2_rank_condition}
\end{equation}
For \(D\neq0\), no branches intersect at a finite point. Instead, the
nodal curve becomes a pair of parallel lines, marking the transition
between a closed conic and a hyperbola-type curve. In the projective
description, the parallel lines meet at infinity. Indeed, homogenizing
the polynomial gives
\[
F(X,Y,Z)=AX^2+BXY+CY^2+DZ^2,
\]
whose degeneracy condition is
\begin{equation}
D\det Q=0
\qquad\Longleftrightarrow\qquad
(a+c)(2ac-b^2)=0 .
\label{eq:N2_projective_discriminant}
\end{equation}
The first factor describes a finite crossing, whereas the second
describes the parallel-line transition.

\begin{figure}[h]
    \centering
    \includegraphics[width=10.0cm]{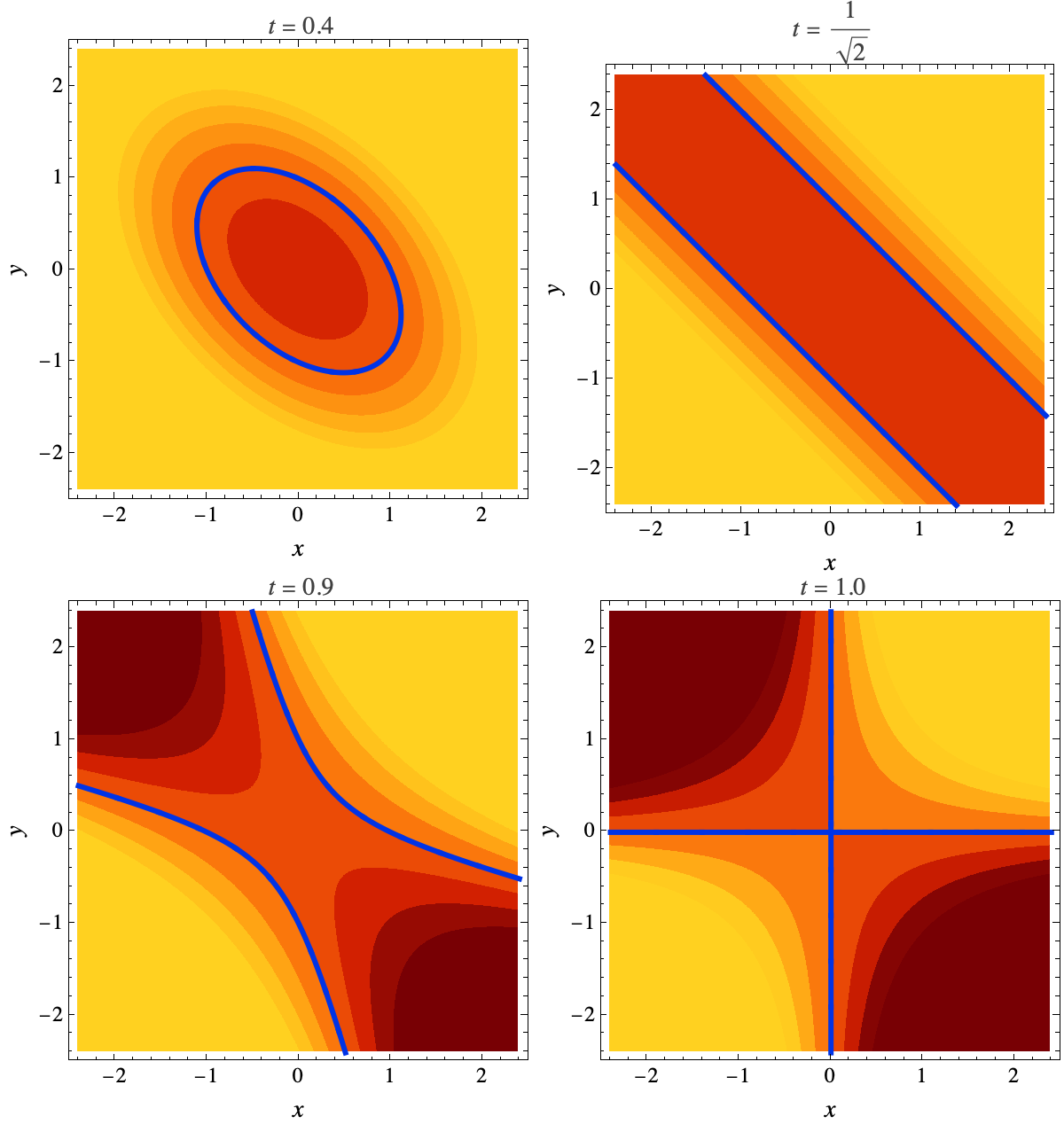}
    \caption{Evolution of the \(N=2\) nodal pattern at \(\alpha=1\) along
    \(a=c=\sqrt{(1-t^2)/2}\), \(b=t\). Top left: closed conic at
    \(t=0.4\). Top right: parallel-line configuration at
    \(t=1/\sqrt{2}\). Bottom left: hyperbola-type curve at \(t=0.9\).
    Bottom right: coordinate cross at \(t=1\).}
    \label{fig:nodal_domains}
\end{figure}

The sign of \(\det Q\) now has a direct geometric interpretation:
\(\det Q>0\) gives a closed ellipse-type curve, \(\det Q<0\) gives an
open hyperbola-type curve, and \(\det Q=0\) gives the intermediate
parallel-line configuration. To follow this transition explicitly, we
choose
\begin{equation}
a=c=\sqrt{\frac{1-t^2}{2}},
\qquad
b=t,
\qquad
t\in[0,1].
\label{eq:N2_path}
\end{equation}
Along this path, the rank loss occurs at \(t^2=1/2\), and the nodal
pattern evolves as follows:
\begin{itemize}
\item \(t^2<1/2\): a closed conic enclosing one of two nodal domains;
\item \(t^2=1/2\): two parallel lines dividing the plane into three domains;
\item \(t^2>1/2\): an open hyperbola-type curve with three domains;
\item \(t=1\): the coordinate cross \(P_2\propto xy\), with four domains.
\end{itemize}
In summary, \(t=1/\sqrt{2}\) is the closed-to-open transition, mediated by two
parallel lines, whereas the endpoint \(t=1\) produces a genuine
finite crossing at the origin. The corresponding evolution of the nodal geometry is displayed in
Fig.~\ref{fig:nodal_domains}.

\subsection{Shannon entropy and smoothness}
\label{subsec:N2_Shannon_smooth}

For (\ref{eq:N2_superposition}), the configuration-space probability density is given by

\begin{equation}
\rho_2(x,y)=\frac{\alpha}{\pi}\,e^{-\alpha(x^2+y^2)}\,P_2(x,y)^2,
\label{eq:N2_rho_explicit}
\end{equation}
and the corresponding entropy $S_r$ (\ref{eq:entropy_2d}) reads

\begin{equation}
 S_r=-\ln\!\left(\frac{\alpha}{\pi}\right)
+\alpha\langle r^2\rangle
-2\langle \ln|P_2|\rangle,
\label{eq:N2_entropy_decomp}   
\end{equation}
where $\langle\cdot\rangle$ denotes expectation with respect to $\rho_2$.
The quadratic moment $\alpha\langle r^2\rangle$ is fixed by the shell index.
From the virial theorem $\langle T\rangle=\langle V\rangle=E_N/2$ with
$E_N$ from~(\ref{eq:EN_def}), and
for any state in the $N$th shell,
\begin{equation}
\frac{1}{2}m\omega^2\langle r^2\rangle=\frac{E_N}{2}
\quad\Rightarrow\quad
\alpha\,\langle r^2\rangle=N+1.
\label{eq:r2_shell_identity}
\end{equation}
For $N=2$,
\begin{equation}
\alpha\,\langle r^2\rangle=3,
\end{equation}
and all coefficient dependence of $S_r$ is therefore carried by
the last term $\langle\ln|P_2|\rangle$.

\paragraph*{Smoothness across the rank-degenerate conic transition $\det Q=0$.}

Although the nodal curve changes topology at $\det Q=0$, the entropy $S_r$
does not develop a singularity there. The zero set has measure zero in
$\mathbb{R}^2$; near a regular zero, $P_2$ vanishes linearly,
while near an isolated singular zero $\mathbf r_c$,
\[
P_2(\mathbf r_c+\boldsymbol{\xi})
=
\frac{1}{2}\boldsymbol{\xi}^{T}
\left[\nabla^2 P_2(\mathbf r_c)\right]
\boldsymbol{\xi},
\]
since $P_2(\mathbf r_c)=0$ and $\nabla P_2(\mathbf r_c)=0$. Because \(\rho_2\propto P_2^2 e^{-\alpha r^2}\), this produces only an integrable logarithmic contribution to \(\langle\ln|P_2|\rangle\), which therefore remains finite and continuous across the transition.

\paragraph*{Analytic checkpoints.}
Two limiting cases provide useful analytic checks.

\emph{(i) Product state $\Phi_{11}$}. At the endpoint $t=1$ of the symmetric 
path~(\ref{eq:N2_path}), the state is proportional to
$\Phi_{11}=\phi_1(x)\phi_1(y)$, so the density factorizes:
\begin{equation}
\rho_1(x)=|\phi_1(x)|^2
=
\sqrt{\frac{\alpha}{\pi}}\,(2\alpha x^2)e^{-\alpha x^2}.
\end{equation}
Hence $S_r=2S_{1\mathrm D,n=1}$, with
\begin{equation}
S_{1\mathrm D,n=1}
=
\frac{1}{2}\ln\!\left(\frac{\pi}{\alpha}\right)
+\gamma+\ln 2-\frac{1}{2},
\end{equation}
so
\begin{equation}
S_r\big|_{\Phi_{11}}
=
\ln\!\left(\frac{\pi}{\alpha}\right)+2\gamma+2\ln 2-1.
\label{eq:S_N2_Phi11_closed}
\end{equation}
This is consistent with $I(x;y)=0$.

\emph{(ii) Quadratic-moment identity.}
Equation~\eqref{eq:r2_shell_identity} supplies a check for all coefficient
choices: any numerical evaluation of $S_r$ in the $N=2$ shell must reproduce
$\alpha\langle r^2\rangle=3$.

Thus the conic degeneracy governs the nodal transition along this path, while
the global Shannon entropy remains a smooth function of the coefficients.

\subsection{Entropic sum}
\label{subsec:N2_entropic_sum}

\begin{figure}[t]
    \centering
    \includegraphics[width=0.6\columnwidth]{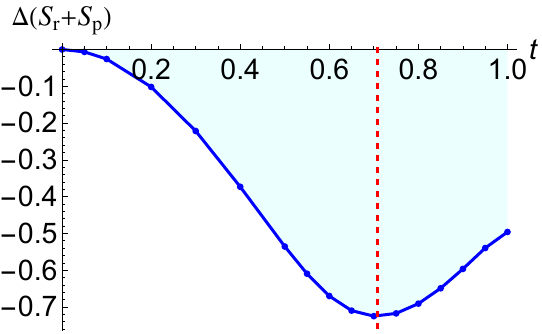}
    \caption{Shifted entropic uncertainty sum
    $\Delta(S_r+S_p)=\bigl(S_r(t)+S_p(t)\bigr)-\bigl(S_r(0)+S_p(0)\bigr)$
    along the symmetric path in the $N=2$ shell. The curve remains smooth near the rank-degenerate conic transition
\(t=1/\sqrt2\) (vertical dotted red line).}
    \label{fig:entropic_sum_N2}
\end{figure}

For fixed-shell oscillator states, the two-dimensional Fourier transform acts
diagonally on the Cartesian basis, and every basis vector in the $N=2$ shell
acquires the same phase factor $(-i)^2=-1$. Consequently, the momentum-space wave 
function $\tilde{\psi}_2(p_x,p_y)$ is given by 
\begin{equation}
\tilde{\psi}_2(p_x,p_y)=
-\frac{1}{m\omega}\,
\psi_2\!\left(\frac{p_x}{m\omega},\frac{p_y}{m\omega}\right).
\end{equation}
Thus
\begin{equation}
\tilde{\rho}_2(p_x,p_y)=
\frac{1}{(m\omega)^2}\,
\rho_2\!\left(\frac{p_x}{m\omega},\frac{p_y}{m\omega}\right),
\end{equation}
and a change of variables yields
\begin{equation}
S_p=S_r+2\ln(m\omega).
\end{equation}
Hence
\begin{equation}
S_r+S_p=2S_r+2\ln(m\omega).
\end{equation}
Hence, the entropic sum inherits the same regularity as $S_r$ and remains
nonsingular across $\det Q=0$ (see Fig.~\ref{fig:entropic_sum_N2}). 
In the dimensionless convention
$m=\omega=\hbar=1$, one has simply $S_p=S_r$ and $S_r+S_p=2S_r$.

\subsection{Nodal-domain entropy and mutual information}
\label{subsec:N2_Sdom_MI}

\begin{figure}[h]
    \centering
    \includegraphics[width=0.6\columnwidth]{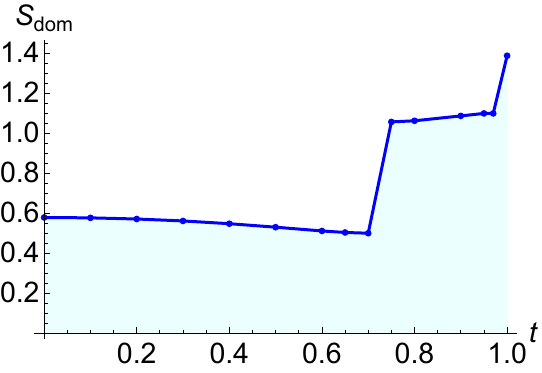}
    \caption{Nodal-domain entropy $S_{\mathrm{dom}}$ at $\alpha=1$ as a function
    of the mixing parameter $t$. Lines are guides to the eye; $S_{\mathrm{dom}}$ is discontinuous at topology-changing points.}
    \label{fig:sdom_vs_t}
\end{figure}

\begin{figure}[t]
    \centering
    \includegraphics[width=0.6\columnwidth]{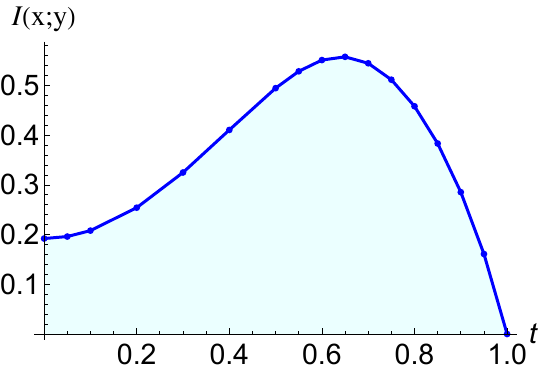}
    \caption{Mutual information $I(x;y)$ versus $t$ at $\alpha=1$ along the
    symmetric path $a=c=\sqrt{(1-t^2)/2}$, $b=t$ in the $N=2$ shell.}
    \label{fig:Ixy}
\end{figure}

Let $\{\Omega_k\}$ be the connected components of
$\mathbb{R}^2\setminus\{P_2=0\}$. Their probabilities $p_k$ (\ref{eq:pk_def}) define $S_{\mathrm{dom}}$ (\ref{eq:Sdom_def}).
For representative numerical calculations, we set $\alpha=1$ (see Fig.~\ref{fig:sdom_vs_t}). 
The point \(t=1/\sqrt{2}\) is a rank-degenerate conic with two parallel
lines, whereas the endpoint \(t=1\) is a finite-affine singularity, where
\(P_2\propto xy\) (see below). We also track the mutual information $I(x;y)$ from
Eq.~\eqref{eq:mutual_information} (see Fig.~\ref{fig:Ixy}).

Two exact checkpoints are especially useful.

\emph{(i) Radial case $t=0$.}
Here $P_2\propto (\alpha r^2-1)$, so the nodal set is the circle
$r=1/\sqrt{\alpha}$, and
\begin{equation}
\begin{aligned}
p_{\mathrm{in}}
&=
\int_{r<1/\sqrt{\alpha}}\rho_2\,dxdy
=
\int_0^1 e^{-s}(s-1)^2\,ds
=
1-\frac{2}{e}\quad ,\qquad
p_{\mathrm{out}}
=\frac{2}{e}.
\end{aligned}
\label{eq:N2_pin_exact}
\end{equation}
Hence
\[
S_{\mathrm{dom}}(0)
=
-p_{\mathrm{in}}\ln p_{\mathrm{in}}
-p_{\mathrm{out}}\ln p_{\mathrm{out}}=0.5774.
\]

\emph{(ii) Endpoint $t=1$.}
By symmetry, the four nodal domains have equal mass $p_k=\tfrac14$, so
\begin{equation}
S_{\mathrm{dom}}(1)=\ln 4,
\qquad
I(x;y)=0,
\end{equation}
since $\rho_2(x,y)\propto x^2e^{-\alpha x^2}y^2e^{-\alpha y^2}$ factorizes.

As expected, \(S_r\) and \(S_r+S_p\) (Fig.~\ref{fig:entropic_sum_N2}) vary smoothly across the
rank-degenerate conic transition, whereas \(S_{\mathrm{dom}}\) (see Fig.~\ref{fig:sdom_vs_t})
responds directly to the changing nodal partition.
The mutual information displayed in Fig.~\ref{fig:Ixy} also changes substantially along this path, reflecting
the correlation restructuring induced by the $xy$ mixing term.

\section{The $N=3$ shell: cubic nodal curves and enhanced restructuring}
\label{secN3}

The $N=3$ shell is fourfold degenerate, with Cartesian basis
\[
\Phi_{30},\;\Phi_{21},\;\Phi_{12},\;\Phi_{03}.
\]
A general real state is
\begin{eqnarray}
\psi_3(x,y)=a\,\Phi_{30}(x,y)\ + \ b\,\Phi_{21}(x,y)\ + \ c\,\Phi_{12}(x,y)\ + \ d\,\Phi_{03}(x,y)\nonumber,
\end{eqnarray}
with
\[
a^2+b^2+c^2+d^2=1.
\]
Using the Hermite representation of the one-dimensional oscillator eigenfunctions, one again obtains the Gaussian--polynomial form $\psi_3(x,y)=e^{-\alpha(x^2+y^2)/2}P_3(x,y)$,
where \(P_3(x,y)\) is a Hermite-constrained real cubic,
\[
P_3(x,y)=
A\,x^3+B\,x^2y+C\,xy^2+D\,y^3+E\,x+F\,y .
\]
Here the coefficients \((A,\ldots,F)\) are linear functions of the shell amplitudes \((a,b,c,d)\) and depend on \(\alpha\). In particular, the lower-order coefficients \(E\) and \(F\) are not independent of the cubic part; they arise from the lower-order terms in the Hermite polynomials.

Again, the nodal set is a real cubic curve. Unlike the linear and quadratic shells, the cubic shell admits a richer geometry of branches and asymptotic directions. Finite-affine restructuring is detected by the singularity conditions (\ref{eq:bifurcation_condition}). Equivalently, the critical point \((x_c,y_c)\) must satisfy
\begin{equation}
\begin{cases}
A x^3 + B x^2y + C xy^2 + D y^3 + E x + F y =0,\\[4pt]
3A x^2 + 2B xy + C y^2 + E =0,\\[4pt]
B x^2 + 2C xy + 3D y^2 + F =0 .
\end{cases}
\label{eq:N3_singularity_equations}
\end{equation}
Eliminating $(x,y)$ from Eq.~\eqref{eq:N3_singularity_equations}
gives an algebraic condition on the cubic coefficients, which we denote by
\begin{equation}
\Delta^{\mathrm{fin}}_3(A,B,C,D,E,F)=0 .
\label{eq:N3_finite_discriminant}
\end{equation}
Here $\Delta^{\mathrm{fin}}_3$ denotes the finite-affine
elimination condition associated with the simultaneous equations
$P_3=\partial_xP_3=\partial_yP_3=0$. Its vanishing identifies coefficient
values at which a finite singularity can occur; a physical nodal singularity
additionally requires a real common solution $(x_c,y_c)$. A corresponding finite-affine resultant, providing an
explicit necessary algebraic condition, is given in the Supplemental
Material.
Finite singularities, however, do not exhaust the relevant restructuring mechanisms. The homogeneous degree-three part of \(P_3\) controls the asymptotic directions of the cubic. Repeated real asymptotic directions define a projective, or asymptotic, degeneracy and can reorganize the arrangement of unbounded branches without producing a finite-affine singularity. Thus the cubic shell naturally separates finite-affine degeneracies from projective ones.

The same organizing principle as for \(N=2\) persists, but
with a richer geometric repertoire. The nodal-domain entropy \(S_{\mathrm{dom}}\) is expected to respond most directly to changes in
the nodal partition, \(I(x;y)\) tracks the induced Cartesian correlations, and the global differential entropies remain less
sensitive because the nodal set has zero measure with respect to the Gaussian-weighted density.

\subsection{Shannon entropy and entropic sum}

A closed form for \(S_r\) is not available throughout the full
\(N=3\) coefficient space, but the regularity mechanism is the same as
in the quadratic shell.  The nodal set \(\{P_3=0\}\) has measure zero
in \(\mathbb{R}^2\).  Near a regular zero, \(P_3\) vanishes linearly,
so \(\ln |P_3|\) is locally integrable against the Gaussian-weighted
density.  At an isolated finite singularity, the logarithmic singularity
is still integrable.  Thus, the configuration-space entropy $S_r$ (\ref{eq:entropy_2d}) remains finite across the cubic families considered below.

What changes across degeneracy strata is not the integrability of
\(S_r\), but the nodal partition itself.  Consequently, \(S_r\) and the
entropic uncertainty sum are expected to be much less sensitive to
finite-affine or projective restructuring than \(S_{\mathrm{dom}}\).

Any nonsmooth behavior of \(S_r\) specifically associated with a
topology-changing nodal event can occur only at an algebraic degeneracy;
however, the global entropy need not display a sharp signature there.

\subsection{Nodal-domain entropy and mutual information: explicit $N=3$ results}

We probe cubic restructuring with the three-state path
\begin{equation}
\psi_3(x,y)
=
a\,\Phi_{30}(x,y)+b\,\Phi_{21}(x,y)+c\,\Phi_{12}(x,y),
\label{eq:N3_three_state}
\end{equation}
where $a^2+b^2+c^2=1$ and the parametrization (\ref{eq:N2_path}) is adopted. This restricted path connects an edge-dominated cubic at $t=0$ to the separable
state $\Phi_{21}$ at $t=1$. Since $\Phi_{03}$ is absent, the family is not
generically invariant under $x\leftrightarrow y$.

Along~\eqref{eq:N2_path}, the polynomial factor has the form
\begin{equation}
P_3(x,y;t)
=
A_1(t)x^3
+A_2(t)x^2y
+A_3(t)xy^2
+B_1(t)x
+B_2(t)y ,
\label{eq:N3_richer_poly}
\end{equation}
with smooth coefficients depending on $t$ and $\alpha$. The lower-order Hermite
terms $B_1x+B_2y$ are essential because they move the critical points and enter
the finite-singularity test.

\begin{figure}[t]
\centering
\includegraphics[width=0.9\linewidth]{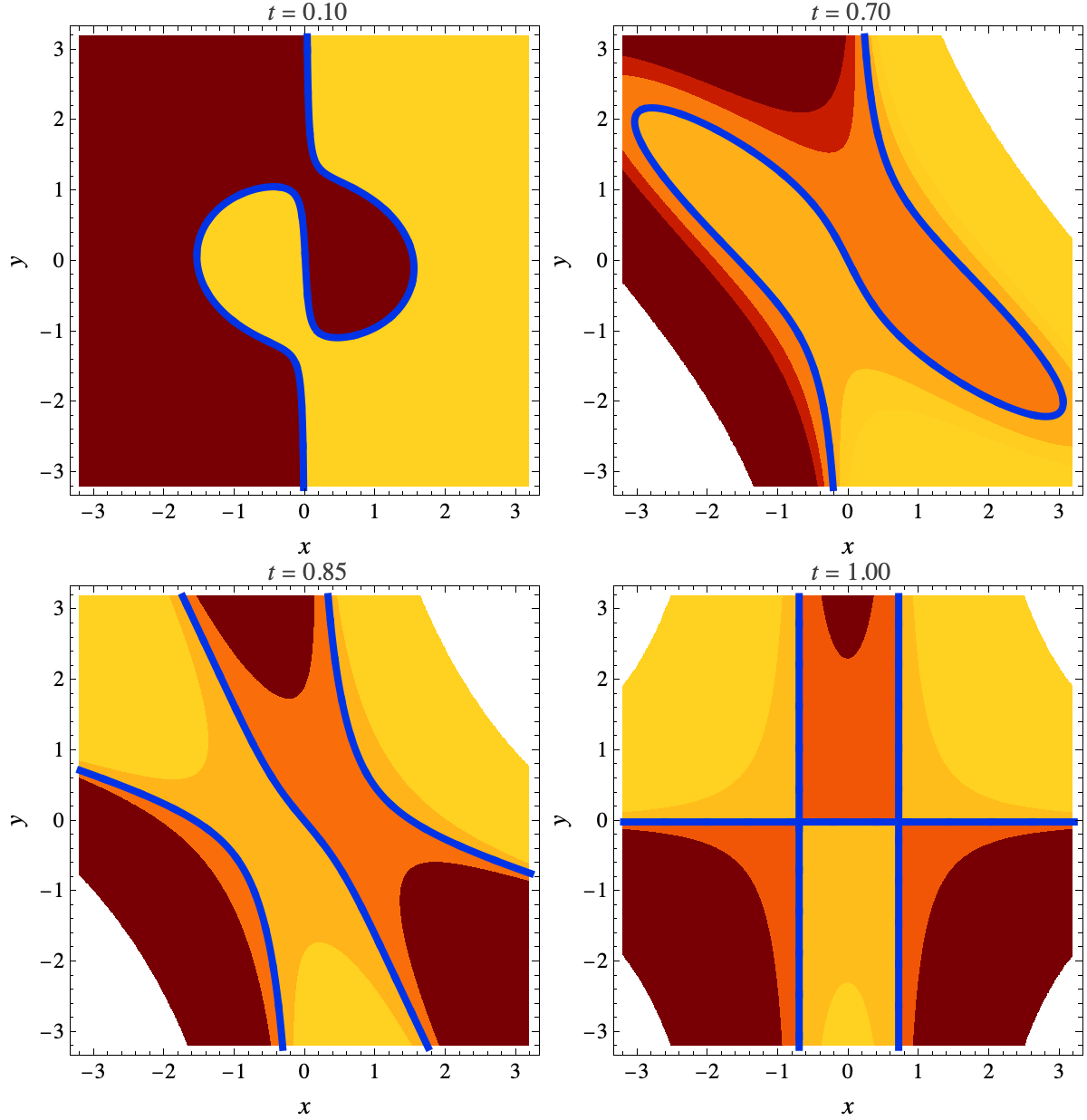}
\caption{
Representative nodal geometries in the \(N=3\) shell at \(\alpha=1\)
along the path~\eqref{eq:N2_path}.
Top left: \(t=0.10\), illustrating the cubic-loop regime.
Top right: \(t=0.70\), near the close-branch regime where smooth nodal branches
approach each other without forming an exact finite-affine singularity.
Bottom left: \(t=0.85\), beyond the close-branch regime, showing a different
nodal arrangement.
Bottom right: \(t=1\), corresponding to the pure \(\Phi_{21}\) state with
nodal set \(y(2\alpha x^2-1)=0\), i.e. the line \(y=0\) together with the two
vertical lines \(x=\pm(2\alpha)^{-1/2}\), which partition the plane into six
nodal domains.
}
\label{fig:N3_nodal}
\end{figure}

We test the relevant conditions (\ref{eq:bifurcation_condition}) with
\begin{equation}
\Delta_{\mathrm{crit}}(t)
=
\min_{\nabla P_3(x_c,y_c;t)=0}
\frac{|P_3(x_c,y_c;t)|}{\|P_3(\cdot,\cdot;t)\|_G},
\end{equation}
where the minimum is taken over the real critical points of
$P_3$, and the Gaussian norm is defined by
\[
\|P_3(\cdot,\cdot;t)\|_G
=
\left[
\iint_{\mathbb R^2}
e^{-\alpha(x^2+y^2)}
P_3(x,y;t)^2\,dx\,dy
\right]^{1/2}.
\]
The condition $\Delta_{\mathrm{crit}}(t)=0$ is equivalent to a real finite-affine singularity.
For $0<t<1$ we find $\Delta_{\mathrm{crit}}(t)>0$, so the close approach of
branches shown in Fig.~\ref{fig:N3_nodal} is not a cusp, crossing, or finite
tangency.

The relevant interior event is projective. Up to a common nonzero factor, the
leading cubic coefficients are
\[
A=\frac{2}{\sqrt3}\alpha^{3/2}a(t),
\qquad
B=2\alpha^{3/2}t,
\qquad
C=2\alpha^{3/2}a(t),
\qquad
D=0 .
\]
The discriminant of the leading binary cubic is
\[
\Delta_\infty=C^2(B^2-4AC),
\]
whose nontrivial zero is
\begin{equation}
t_\infty=\sqrt{4-2\sqrt3}\simeq0.732 .
\end{equation}
At \(t=t_\infty\), two asymptotic directions of the cubic coalesce even though the affine curve remains nonsingular. The close encounter of the branches in Fig.~\ref{fig:N3_nodal} is controlled by the geometry at infinity rather than by a cusp, crossing, or tangency at any finite point.

The endpoints are singular in a different sense. At $t=0$, $a=c=1/\sqrt2$ and
$b=0$, giving
\[
P_3(x,y;0)
\propto
x\left[
\frac{2}{\sqrt3}\alpha x^2
+
2\alpha y^2
-
(\sqrt3+1)
\right].
\]
The nodal set is the union of the line $x=0$ and the ellipse
\[
\frac{2}{\sqrt3}\alpha x^2
+
2\alpha y^2
=
\sqrt3+1 .
\]
For small positive $t$, this line--ellipse configuration unfolds into the
looped cubic shown in Fig.~\ref{fig:N3_nodal}. At $t=1$,
\[
P_3(x,y;1)\propto y(2\alpha x^2-1),
\]
so the nodal set consists of $y=0$ and $x=\pm(2\alpha)^{-1/2}$, which partition
the plane into six domains.

\begin{figure}[t]
    \centering
    \includegraphics[width=0.6\columnwidth]{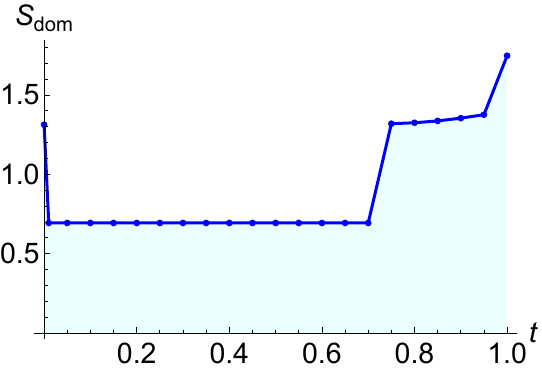}
    \caption{
    Nodal-domain entropy \(S_{\rm dom}\) at \(\alpha=1\) along the
\(N=3\) path \eqref{eq:N2_path}. The sharp change occurs near
\(t_\infty=\sqrt{4-2\sqrt3}\simeq0.732\), where the leading cubic reaches the
projective discriminant. The finite-affine diagnostic satisfies
\(\Delta_{\rm crit}>0\) for \(0<t<1\), so the close-branch regime is not a
finite cusp, crossing, or tangency. Lines are guides to the eye; $S_{\mathrm{dom}}$ is discontinuous at topology-changing points.
    }
    \label{fig:SdomN3}
\end{figure}

\begin{figure}[t]
    \centering
    \includegraphics[width=0.6\columnwidth]{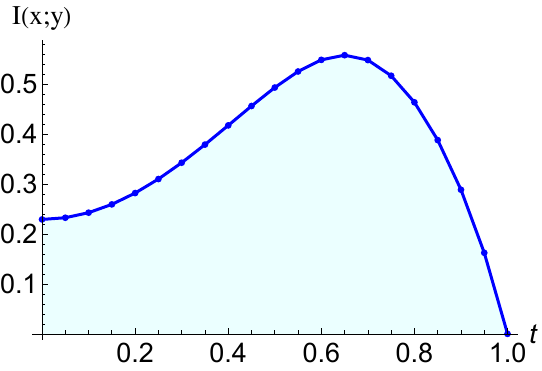}
    \caption{
    Mutual information \(I(x;y)\) versus \(t\) at \(\alpha=1\) along the
    three-state path \eqref{eq:N2_path} in the \(N=3\) shell. The curve
    rises as the cubic nodal geometry becomes more strongly correlated, reaches
    a maximum near the projective close-branch regime, and decreases to zero at
    \(t=1\), where the state reduces to the separable basis state \(\Phi_{21}\).
    }
    \label{fig:IxyN3}
\end{figure}

The entropy and correlation diagnostics are shown in
Figs.~\ref{fig:SdomN3} and \ref{fig:IxyN3}. The sharp change in
\(S_{\rm dom}\) occurs near
\(t_\infty=\sqrt{4-2\sqrt3}\simeq0.732\), the projective-discriminant value.
The interior values \(0<t<1\) of \(S_{\rm dom}\) were computed from the
connected components of the sign field of \(P_3\), while the endpoint values
were assigned from the analytic factorizations above. In all numerical panels we set \(\alpha=1\).

Thus the \(N=3\) path separates endpoint degeneracies at \(t=0,1\), a smooth
affine curve for \(0<t<1\), and a projective restructuring at \(t_\infty\). The
strongest response of both \(S_{\rm dom}\) and \(I(x;y)\) occurs near the
projective close-branch regime, while \(S_r+S_p\) remains comparatively
insensitive because the nodal set has zero measure.

\section{A representative family and asymptotic structure for general $N$}
\label{sec:generalN}

To extend the low-shell analysis without introducing the full
$(N+1)$-parameter coefficient space, we consider, for \(N\ge2\),
the one-parameter family defined by
\begin{equation}
\begin{aligned}
\psi_N^{(t)}
&=
\sqrt{1-t^2}\,\frac{\Phi_{N,0}(x,y)+\Phi_{0,N}(x,y)}{\sqrt2}
+t\,\Phi_{n_+,n_-}(x,y),\qquad
 0\le t\le 1.
\end{aligned}
\label{eq:general-family}
\end{equation}
here $n_-=\Big\lfloor\frac N2\Big\rfloor,
n_+=\Big\lceil\frac N2\Big\rceil$. This path interpolates between an edge-dominated superposition at $t=0$ and the
separable product state at $t=1$. It is a representative path of the fixed shell, not a unique canonical path through the full $(N+1)$-dimensional
coefficient space. In particular, for $N=3$ it is not identical to
the restricted three-state family studied in Sec.~\ref{secN3}, although it
implements the same interpolation between edge-dominated
superpositions and a separable Cartesian product state.

With $\xi=\sqrt{\alpha}\,x$ and $\eta=\sqrt{\alpha}\,y$, one has the
Gaussian--polynomial form
\begin{equation}
\psi_N^{(t)}(x,y)=\mathcal N_N\, e^{-\alpha\,r^2/2}\,Q_N^{(t)}(\xi,\eta),
\label{eq:general-family-gaussian}
\end{equation}
where $\mathcal N_N\neq 0$ and
\begin{eqnarray}
Q_N^{(t)}(\xi,\eta) &=& \frac{\sqrt{1-t^2}}{\sqrt2}\,\bigl[H_N(\xi)+H_N(\eta)\bigr]\
                    + \ t\,\sqrt{\binom{N}{n_-}}\,H_{n_+}(\xi)H_{n_-}(\eta). \nonumber  
\label{eq:general-Q}                    
\end{eqnarray}

Naturally, the nodal curve is given by $Q_N^{(t)}(\xi,\eta)=0.$ At the endpoint $t=1$,
\begin{equation}
\psi_N^{(1)}(x,y)=\Phi_{n_+,n_-}(x,y)=\phi_{n_+}(x)\phi_{n_-}(y),
\label{eq:general-endpoint-state}
\end{equation}
so the nodal set becomes a rectangular grid of Hermite zeros and the plane is
partitioned into
\begin{eqnarray}
\nu_N(1)&=&(n_++1)(n_-+1)\\
&=&
\begin{cases}
(\ell+1)^2, & \textrm{if}\,\,\,\, N=2\ell,\\[1mm]
(\ell+1)(\ell+2), & \textrm{if}\,\,\,\, N=2\ell+1, \nonumber
\end{cases}
\label{eq:general-endpoint-count}
\end{eqnarray}
where $\ell$ is an integer. 
Since the density factorizes at $t=1$, so do the nodal-domain weights, and hence
\begin{equation}
S_{\mathrm{dom}}(1)=S_{\mathrm{dom}}^{1\mathrm D}(n_+)+S_{\mathrm{dom}}^{1\mathrm D}(n_-),
\qquad
I(x;y)\big|_{t=1}=0.
\label{eq:general-endpoint-factorization}
\end{equation}
The derivation is given in the Supplemental Material.

Similarly to (\ref{eq:bifurcation_condition}), finite singular events along the path are determined by $Q_N^{(t)}(\xi_c,\eta_c)=0$ and $\nabla Q_N^{(t)}(\xi_c,\eta_c)=0$. Using $H_n'(z)=2n\,H_{n-1}(z)$, this becomes
\begin{align}
&\frac{\sqrt{1-t^2}}{\sqrt2}\,\bigl[H_N(\xi)+H_N(\eta)\bigr]
+t\sqrt{\binom{N}{n_-}}\,H_{n_+}(\xi)H_{n_-}(\eta)=0,
\label{eq:general-singularity-a}\\
&\frac{\sqrt{1-t^2}}{\sqrt2}\,H_{N-1}(\xi)
+t\sqrt{\binom{N}{n_-}}\frac{n_+}{N}\,H_{n_+-1}(\xi)H_{n_-}(\eta)=0,
\label{eq:general-singularity-b}\\
&\frac{\sqrt{1-t^2}}{\sqrt2}\,H_{N-1}(\eta)
+t\sqrt{\binom{N}{n_-}}\frac{n_-}{N}\,H_{n_+}(\xi)H_{n_--1}(\eta)=0.
\label{eq:general-singularity-c}
\end{align}
These equations describe the singular locus even when an explicit discriminant
elimination is not available.

The family also makes the large-distance structure transparent. Since
\[
H_n(z)=2^n z^n+\mathcal O(z^{n-2}),
\]
the leading homogeneous part of $Q_N^{(t)}$ is
\begin{equation}
Q_{N,\mathrm{top}}^{(t)}(\xi,\eta)
=
2^N\left[
\frac{\sqrt{1-t^2}}{\sqrt2}\,(\xi^N+\eta^N)
+t\sqrt{\binom{N}{n_-}}\,\xi^{n_+}\eta^{n_-}
\right].
\label{eq:general-top}
\end{equation}
Writing $(\xi,\eta)=(r\cos\theta,r\sin\theta)$ gives
\begin{equation}
Q_{N,\mathrm{top}}^{(t)}=2^N r^N f_{N,t}(\theta),
\label{eq:general-angular-form}
\end{equation}
with
\begin{equation}
f_{N,t}(\theta)
=
\frac{\sqrt{1-t^2}}{\sqrt2}\bigl(\cos^N\theta+\sin^N\theta\bigr)
+t\sqrt{\binom{N}{n_-}}\cos^{n_+}\theta\,\sin^{n_-}\theta.
\label{eq:general-angular}
\end{equation}
Repeated zeros of \(f_{N,t}\), characterized by
\[
f_{N,t}(\theta)=0,\qquad f'_{N,t}(\theta)=0,
\]
give the corresponding projective or asymptotic degeneracies. Thus, the asymptotic nodal rays are determined by
\begin{equation}
f_{N,t}(\theta)=0.
\label{eq:general-rays}
\end{equation}

More generally, the large-radius nodal geometry is controlled by the top
homogeneous component: lower-degree Hermite terms affect the curve locally,
whereas the arrangement of ends is fixed by the degree-$N$ part. From the
symmetry viewpoint, this leading term is the highest-degree component of the
spin-$j=N/2$ $SU(2)$ multiplet and therefore encodes the asymptotic
interference pattern of the superposition.

Equations~\eqref{eq:general-singularity-a}--\eqref{eq:general-rays} extend the
low-shell analysis to arbitrary shell number \(N\ge2\). This family connects the low-shell examples with arbitrary \(N\): its separable endpoint is known exactly, while its finite and asymptotic degeneracies can be followed directly along a single path in coefficient space.

\section{Physical implementation}
\label{sec:physical_implementation}

The fixed-shell structure analyzed above is not tied to a particular
realization of the oscillator. It can be implemented in systems where a
two-dimensional harmonic mode structure is available and where coherent
superpositions within a fixed shell can be prepared, controlled, and
reconstructed. Two natural platforms are Hermite--Gaussian structured
light and two-dimensional trapped motional states. In both cases the
relevant observables are the same: the near-field or position-space
density gives the nodal-domain weights and Cartesian mutual information,
while the far-field or momentum-space density gives the momentum entropy
entering \(S_r+S_p\).

\subsection{Structured-light realization}

In paraxial optics, Hermite--Gaussian modes are transverse oscillator
eigenmodes. A coherent superposition with fixed transverse order
\(N=n_x+n_y\) and real relative phases at a chosen propagation plane has,
after transverse rescaling, the Gaussian--polynomial form
\begin{equation}
u_N(x,y)=e^{-r^2/w^2}P_N(x/w,y/w),
\label{eq:optical-gaussian-polynomial}
\end{equation}
up to an overall scale and phase, where \(r^2=x^2+y^2\) and \(w=w(z)\)
is the Gaussian beam radius at the selected propagation plane. With the
convention used in Eq.~\eqref{eq:optical-gaussian-polynomial}, \(w\) is
the radius at which the field amplitude decreases to \(1/e\) of its
on-axis value, equivalently the intensity decreases to \(1/e^2\). If the
selected plane is the beam-waist plane, then \(w=w_0\), where \(w_0\)
denotes the waist radius. Numerical factors associated with the standard
Hermite--Gaussian convention, such as \(\sqrt{2}\) in the polynomial
arguments, are absorbed into the definition of \(P_N\) and the
transverse rescaling.

Since the Gaussian envelope is strictly positive, the dark curves of the
optical field are determined by $P_N(x/w,y/w)=0$. So, the nodal algebraic curves studied in the previous sections are
directly realized as zero-amplitude curves of a real-phase
Hermite--Gaussian field. The nodal curves themselves are obtained from
the intensity zeros, while the restriction to a real field can be
verified or enforced by phase control or interferometric reconstruction.

Recent structured-light platforms allow programmable control of
transverse amplitude and phase, including higher-order free-space modes,
making coefficient-space paths of the type considered here
experimentally accessible~\cite{Forbes2021,Butow2024}. Given a measured
transverse intensity \(I_{\rm opt}(x,y)\), one forms the normalized
density
\begin{equation}
\rho(x,y)=
\frac{I_{\rm opt}(x,y)}
{\iint_{\mathbb R^2} I_{\rm opt}(x,y)\,dx\,dy}.
\label{eq:optical-density}
\end{equation}
The nodal-domain weights $p_k$ (\ref{eq:pk_def}) are then obtained by integrating \(\rho\) over
the connected bright regions \(\Omega_k\) separated by the dark curves. The Cartesian mutual information follows from the measured marginals (\ref{eq:marginals}),
through \(I(x;y)=S_x+S_y-S_r\). The momentum-space entropy \(S_p\) (\ref{eq:entropy_p}) is
accessed from the far-field intensity, which is proportional to the
squared modulus of the transverse Fourier transform of \(u_N(x,y)\).
Hence, structured light provides a direct optical route to all three
diagnostics:
\[
S_{\rm dom},\qquad I(x;y),\qquad S_r+S_p .
\]
A sharp response of \(S_{\rm dom}\) accompanied by a smooth \(S_r+S_p\)
would be the optical signature of nodal restructuring at fixed
transverse order.

\subsection{Trapped motional states}

A complementary realization is provided by the transverse motion of a
single atom or ion in an approximately isotropic two-dimensional harmonic
trap. When the transverse frequencies are equal, the product states
\(\Phi_{n,N-n}(x,y)\) in Eq.~\eqref{eq:cart_basis} have the same energy and
span the degenerate motional subspace
\begin{equation}
{\cal H}_N
=
\mathrm{span}\{\Phi_{n,N-n}:n=0,\ldots,N\}.
\end{equation}
Coherent preparation and control of two-dimensional motional
superpositions have been demonstrated in trapped ions~\cite{Jeon2024}.

Exact degeneracy requires equal transverse frequencies. A small
anisotropy lifts this degeneracy and may be treated as a controlled
perturbation. The nodal analysis is thus most directly applicable
to operations designed to remain within a single nearly isotropic shell,
with population outside the target shell monitored independently in the
oscillator basis.

For a reconstructed real motional wavefunction \(\psi_N(x,y)\), as in
Eq.~\eqref{eq:gauss_poly}, the position distribution determines
\(S_r\), \(S_{\rm dom}\), and \(I(x;y)\), while the momentum distribution
determines \(S_p\). States with the same energy and similar global
delocalization can nevertheless have different nodal patterns. Accordingly, the
number, adjacency, and probability weights of the nodal domains
provide information not contained in the shell population or the global
position--momentum entropy.

\subsection{Nodal verification of fixed-shell operations}

A controlled operation within a degenerate oscillator shell changes the
superposition coefficients while preserving the energy. The resulting
deformation of the nodal pattern provides a state-sensitive
test complementary to energy and shell-population measurements. We
consider operations that preserve real coefficients, for which the nodes
form a single real curve \(P_N(x,y)=0\). Complex superpositions may
instead support phase singularities and require a separate analysis.

Let
\begin{equation}
|n\rangle_N\equiv \Phi_{n,N-n},
\qquad n=0,\ldots,N,
\end{equation}
denote the Cartesian basis of the shell. An ideal operation generates a
path of normalized real coefficient vectors \(c_G(s)\), with
\(0\leq s\leq 1\), and hence a family of Hermite-constrained polynomials
\[
P_{c_G(s)}(x,y)
\]
with nodal curves
\begin{equation}
Z_G(s)=\{(x,y):P_{c_G(s)}(x,y)=0\}.
\label{eq:gate-nodal-curve}
\end{equation}
Thus, states with the same energy can be distinguished through the
geometry and probability distribution associated with \(Z_G(s)\).

The nodal topology can change through two physical mechanisms: branches
may meet at a finite point, or asymptotic nodal directions may merge.
The corresponding coefficient sets are
\begin{equation}
\Sigma_N=\Sigma_N^{\rm aff}\cup\Sigma_N^\infty ,
\end{equation}
where
\begin{equation}
\Sigma_N^{\rm aff}
=
\{c:\exists(x,y),\;P_c=\partial_xP_c=\partial_yP_c=0\}
\label{eq:affine-degeneracy-set}
\end{equation}
describes finite crossings or tangencies, while
\begin{equation}
\begin{aligned}
\Sigma_N^\infty
&=
\{c:\exists\,\theta,\;f_c(\theta)=f_c'(\theta)=0\},
\qquad
f_c(\theta)
=
P_{c,\mathrm{top}}(\cos\theta,\sin\theta)
\end{aligned}
\label{eq:projective-degeneracy-set}
\end{equation}
identifies mergers of asymptotic directions. If \(c_G(s)\) remains a
positive distance from \(\Sigma_N\), the nodal curve deforms smoothly
without reconnection. The number and adjacency of the nodal domains
remain fixed. Let \(\Omega_k^G(s)\) denote the \(k\)th connected nodal domain
associated with \(Z_G(s)\). The corresponding ideal probability weights,
\begin{equation}
p_k^G(s)=
\int_{\Omega_k^G(s)}
|\psi_G(x,y;s)|^2\,dx\,dy,
\label{eq:gate-domain-weights}
\end{equation}
vary smoothly. The predicted nodal response can be summarized
by
\begin{equation}
{\cal F}_G(s)=
\bigl(\tau_G(s),\{p_k^G(s)\},S_{\rm dom}^G(s),I_G(x;y;s)\bigr),
\label{eq:gate-fingerprint}
\end{equation}
where \(\tau_G(s)\) specifies the domain-adjacency pattern.

Experimentally, reconstruction of the state or probability density gives
\[
\widehat{\tau}(s),
\qquad
\{\widehat p_k(s)\},
\qquad
\widehat S_{\rm dom}(s),
\qquad
\widehat I(x;y;s).
\]
An unexpected change in topology along a path designed to avoid
\(\Sigma_N\) directly signals imperfect control. When the measured and
predicted topologies agree, their quantitative difference can be measured
without assigning fixed labels to the domains:
\begin{align}
D_{\rm node}(s)
&=
\min_{\pi}
\Bigg[
\sum_k
\left|\widehat p_k(s)-p^G_{\pi(k)}(s)\right|
+
\lambda_I
\left|\widehat I(x;y;s)-I_G(x;y;s)\right|
\Bigg]. \nonumber
\end{align}
Here \(\pi\) runs over domain relabelings and \(\lambda_I\) sets the
relative weight of the mutual-information contribution. A large
\(D_{\rm node}\), an altered adjacency pattern, or the appearance or
loss of a nodal domain indicates a control error, population leakage
from the target shell, or inaccurate state reconstruction. These effects
need not be visible in the total energy or shell populations, which do
not determine the superposition within a degenerate shell.

For \(N=2\), this test is explicit. Along the path
(\ref{eq:N2_path}), the conic reaches the rank-degeneracy condition
\(b^2=2ac\) at \(t=1/\sqrt{2}\). The nodal curve then changes from a
closed conic to parallel lines and subsequently to a hyperbola-type
curve. Although \(S_r+S_p\) remains smooth, \(S_{\rm dom}\) responds
directly to the accompanying change in the nodal partition. As a result, the nodal
observables can reveal deviations from the intended evolution
that are invisible to energy conservation or global delocalization
measures.

The degeneracy sets thus provide experimentally identifiable landmarks for controlled trajectories within a fixed shell and offer a geometry-based complement to trapped-ion motional-state tomography
\cite{FluehmannHome2020}.

\section{Conclusion}

We have shown that degeneracy allows the nodal geometry of a quantum
state to change without changing its energy. For the two-dimensional
isotropic harmonic oscillator, every real state in a fixed shell can be
written as a Gaussian envelope multiplied by a constrained polynomial,
Eq.~\eqref{eq:gauss_poly}. Because the Gaussian never vanishes, the
observable nodal pattern is determined entirely by the curve
\(P_N(x,y)=0\). Its topology can change when nodal branches meet at a
finite point or when two asymptotic directions merge.

Theorem~\ref{thm:regular-family} identifies the complementary regular
regime. If neither event occurs, the nodal curve can bend and move but
cannot reconnect. Consequently, the number of nodal domains remains
fixed, their probabilities vary smoothly, and \(S_{\rm dom}\) has no
singular behavior. This result separates genuine changes of nodal
connectivity from smooth geometric deformations within the same
energy shell.

The first shells illustrate this mechanism directly. For \(N=1\),
changing the coefficients only rotates a straight nodal line, leaving
two equal-probability domains and \(S_{\rm dom}=\ln 2\). For \(N=2\), a
closed conic becomes an open hyperbola-type curve through an intermediate
pair of parallel lines at \(b^2=2ac\). The nodal-domain entropy responds
clearly to this restructuring, while the global position--momentum
entropy remains smooth. For \(N=3\), the merger of two asymptotic
directions brings smooth cubic branches close together without producing
a finite crossing. This geometry enhances both \(S_{\rm dom}\) and the
Cartesian mutual information \(I(x;y)\).

The diagnostics therefore probe distinct physical properties.
The domain probabilities \(p_k\), defined in Eq.~\eqref{eq:pk_def}, and
their entropy \(S_{\rm dom}\), Eq.~\eqref{eq:Sdom_def}, measure how
probability is redistributed among the regions separated by the nodes.
The mutual information \(I(x;y)\), Eq.~\eqref{eq:mutual_information},
measures correlations between the two Cartesian coordinates, whereas
\(S_r+S_p\) characterizes the overall position--momentum delocalization.
The comparatively strong response of \(S_{\rm dom}\) reflects its direct
dependence on the nodal partition.

These quantities are experimentally accessible. In real-phase
Hermite--Gaussian beams, the nodal curves are the dark lines of the
transverse intensity, and the probabilities \(p_k\) follow by integrating
the intensity over the connected bright regions. Near- and far-field
measurements provide the position- and momentum-space distributions. The
same analysis applies to reconstructed motional states in an
approximately isotropic two-dimensional trap. In either platform, an
unexpected nodal reconnection or a deviation of the measured domain
weights from their predicted values provides a sensitive test of
controlled evolution within a fixed shell, complementary to energy,
shell-population, and state-tomography measurements.

The mechanism is not restricted to the two-dimensional oscillator.
Appendix~B shows that the first three-dimensional shells contain nodal
planes for \(N=1\) and topology-changing quadric surfaces for \(N=2\).
The two-dimensional square-well example presented in the
Supplemental Material demonstrates an additional hard-wall mechanism:
a nodal curve can change connectivity by touching the boundary. These examples indicate that fixed-energy nodal
restructuring is a general consequence of degeneracy rather than a
special property of Gaussian--Hermite states.

A natural extension is the hydrogen atom. In a real spherical-harmonic
basis, a state in a fixed principal shell can be written as
\[
\psi_n(\mathbf r)
=
\sum_{\ell=0}^{n-1}
\sum_{m=-\ell}^{\ell}
c_{\ell m}
R_{n\ell}(r)
Y_{\ell m}^{(\mathrm R)}(\theta,\phi).
\]
Here the nodes are surfaces rather than planar curves, but the same
physical questions remain: when can they reconnect at fixed energy, and
which observables best detect the associated redistribution of
probability? Further directions include complex superpositions with
phase singularities, commensurate anisotropic oscillators, higher
degenerate shells, and semiclassical regimes with increasingly intricate
nodal patterns.

\section*{Data availability statement}

All data that support the findings of this study are included within the article. 
The numerical results reported in the figures were generated from the analytic 
expressions given in the manuscript, appendix and Supplemental Material.

\section*{Acknowledgments}
A.M.E.R. thanks R.~Sagar for insightful discussions.

\appendix

\section{Proof of Theorem~\ref{thm:regular-family}}
\label{proof}

Set
\[
F:J\times\mathbb R^2\to\mathbb R,
\qquad
F(t,x,y)=P_t(x,y).
\]
By assumption (i), $0$ is a regular value of each map $F_t:=F(t,\cdot,\cdot)$,
so $Z_t=F_t^{-1}(0)$ is a smooth one-dimensional submanifold of $\mathbb R^2$.
Indeed, on $Z_t$ one has
$\nabla P_t\neq0$, so the implicit-function theorem gives a smooth local
parametrization of the nodal curve near every finite point
\cite{Hirsch1976}. Moreover,
\[
P_t(r\cos\theta,r\sin\theta)=r^N f_t(\theta)+O(r^{N-1}),
\qquad r\to\infty,
\]
uniformly in $t\in J$. Since the zeros of $f_t$ are simple by assumption (ii),
each asymptotic nodal direction persists under a small change
of $t$ and remains separated from the others. By compactness of $J$, this
separation can be chosen uniformly in $t$. Hence there exists $R>0$ such that
$Z_t\cap\partial B_R$ is transverse for all $t\in J$, with a $t$-independent
number of intersection points. The nodal branches outside
$B_R$ therefore form a fixed number of smooth ends, varying continuously with
$t$, so each $Z_t$ is properly embedded.

Inside $\overline{B_R}$, the set
\[
W:=F^{-1}(0)\cap (J\times\overline{B_R})
\]
is a \(C^2\) submanifold. The projection
\[
\pi:W\longrightarrow J,\qquad
\pi(t,x,y)=t,
\]
is a submersion. To see this, note that at every point of $W$ the spatial
gradient $\nabla P_t$ is nonzero by assumption (i); hence, for any infinitesimal
variation of $t$, the tangency condition to $W$ can be satisfied by an
appropriate variation in $(x,y)$. Since $J\times\overline{B_R}$ is compact,
the restricted projection is proper. The transversality of
$Z_t\cap\partial B_R$ established above ensures that this submersion is
compatible with the boundary.

Since \(F\) is \(C^2\), the projection
\(\pi:W\to J\) is a \(C^2\) proper submersion. By the Ehresmann fibration
theorem for proper submersions \cite{Dundas2018}, it therefore admits a
local \(C^1\) trivialization. Consequently, the portions
$Z_t\cap\overline{B_R}$ form a locally trivial family. Matching this
trivialization to the \(C^1\)-varying exterior ends gives a deformation of
the complete nodal curve. The isotopy-extension theorem then promotes this
deformation to an ambient isotopy of $\mathbb R^2$ \cite{Hirsch1976}, giving
\[
H_t(Z_{t_0})=Z_t
\]
as claimed in Theorem~\ref{thm:regular-family}(a). This implies local constancy of the number
of nodal domains, hence global constancy on the connected interval $J$.

Fix $t_0\in J$. The ambient isotopy provides a consistent
labeling of the nodal domains $\Omega_k(t)$ for $t$ near $t_0$. Their
probability weights are
\[
p_k(t)=\int_{\Omega_k(t)}\rho_t(x,y)\,dx\,dy .
\]
Since the boundaries of the nodal domains move through a $C^1$ family,
the transport theorem gives
\[
\frac{d}{dt}p_k(t)
=
\int_{\Omega_k(t)}\partial_t\rho_t(x,y)\,dx\,dy
+
\int_{\partial\Omega_k(t)}\rho_t\,v_n\,ds ,
\]
where $v_n$ denotes the normal velocity of the moving boundary. By definition, \(\partial\Omega_k(t)\subset Z_t\), and therefore \(\rho_t=|\psi_t|^2=0\) on the nodal boundary. Hence the boundary term vanishes. For unbounded nodal domains, the same result follows by applying
the transport formula to $\Omega_k(t)\cap B_R$ and taking $R\to\infty$;
the contribution from $\partial B_R$ vanishes because of the Gaussian
decay.

Since the coefficients of $P_t$ vary continuously on compact $J$, there exists
$C>0$ such that
\[
\begin{aligned}
0\le \rho_t(x,y)&\le C(1+r^{2N})e^{-\alpha r^2},\\
|\partial_t\rho_t(x,y)|&\le C(1+r^{2N})e^{-\alpha r^2}.
\end{aligned}
\]
The right-hand side is integrable uniformly in $t$, so the
volume integral above is well defined and varies continuously with $t$.
It follows that each $p_k$ is $C^1$.
Since $\Omega_k(t)$ is a nonempty open set and $\rho_t>0$ away from $Z_t$,
one has $p_k(t)>0$.

Because $u\mapsto -u\ln u$ is smooth on $(0,1)$, the entropy
\[
S_{\rm dom}(t)=-\sum_{k=1}^{M}p_k(t)\ln p_k(t)
\]
is $C^1$.

For the configuration-space entropy, let $t_n\to t_0$. The same Gaussian bound
gives an integrable majorant
\[
0\le \rho_t(x,y)\le g(x,y):=C(1+r^{2N})e^{-\alpha r^2}\in L^1(\mathbb R^2),
\]
and pointwise convergence $\rho_t\to\rho_{t_0}$. Using
\[
|u\ln u|\le C'(u+u^{1/2}),
\qquad u\ge 0,
\]
it follows that
\[
|\rho_t\ln\rho_t|\le C''(g+g^{1/2})\in L^1(\mathbb R^2),
\]
so dominated convergence yields continuity of $S_r$.

Finally, all basis states in the fixed $N$ shell acquire the same Fourier phase
$(-i)^N$, so
\[
\widetilde\psi_t(p_x,p_y)=\frac{(-i)^N}{m\omega}\,
\psi_t\!\left(\frac{p_x}{m\omega},\frac{p_y}{m\omega}\right),
\]
hence
\[
\widetilde\rho_t(p_x,p_y)=\frac{1}{(m\omega)^2}\,
\rho_t\!\left(\frac{p_x}{m\omega},\frac{p_y}{m\omega}\right).
\]
A change of variables gives
\[
S_p(t)=S_r(t)+2\ln(m\omega),
\]
so $S_r(t)+S_p(t)$ is continuous.\hfill\(\square\)

\section{First harmonic-oscillator shells in three dimensions}
\label{sec:3d_extension}

For the three-dimensional isotropic harmonic oscillator, the shell quantum number, energy, and degeneracy are
\begin{equation}
N=n_x+n_y+n_z,\qquad
E_N=\hbar\omega\left(N+\frac{3}{2}\right),\qquad
g_N=\binom{N+2}{2}.
\end{equation}
A real state in the \(N\)th shell has the Gaussian--polynomial form
\begin{equation}
\psi_N(\mathbf{x})
=e^{-\alpha r^2/2}P_N(\mathbf{x}),
\qquad
\alpha=\frac{m\omega}{\hbar},
\qquad
r^2=x^2+y^2+z^2,
\end{equation}
so its nodal set is the algebraic surface \(P_N(\mathbf{x})=0\). The nodal-domain entropy carries over directly, with volume integrals replacing area integrals. The ground shell \(N=0\) has no nodes and \(S_{\rm dom}=0\).

\emph{\(N=1\).---}
The first excited shell is threefold degenerate:
\begin{equation}
\psi_1=a\Phi_{100}+b\Phi_{010}+c\Phi_{001},
\qquad
a^2+b^2+c^2=1.
\end{equation}
Apart from an overall nonzero factor,
\begin{equation}
P_1(\mathbf{x})=ax+by+cz.
\end{equation}
Thus every real state has a single nodal plane through the origin. Because
\(\nabla P_1=(a,b,c)\neq0\), coefficient variation only rotates this plane and cannot produce a nodal transition. The plane divides space into two half-spaces of equal probability, giving
\begin{equation}
S_{{\rm dom},N=1}=\ln 2.
\end{equation}

\emph{\(N=2\).---}
The second excited shell is sixfold degenerate. Write
\begin{equation}
\psi_2=
a\Phi_{200}+b\Phi_{020}+c\Phi_{002}
+d\Phi_{110}+e\Phi_{101}+f\Phi_{011},
\end{equation}
with
\begin{equation}
a^2+b^2+c^2+d^2+e^2+f^2=1.
\end{equation}
Its polynomial factor can be expressed as
\begin{equation}
P_2(\mathbf{x})
=\mathbf{x}^{\mathsf T}Q\mathbf{x}
-\frac{\operatorname{tr}Q}{2\alpha},
\qquad
Q=\alpha
\begin{pmatrix}
\sqrt{2}a & d & e\\
d & \sqrt{2}b & f\\
e & f & \sqrt{2}c
\end{pmatrix}.
\label{eq:3d_quadric}
\end{equation}
Hence the nodal surfaces are Hermite-constrained centered quadrics. After a spatial rotation,
\begin{equation}
P_2=
\lambda_1u^2+\lambda_2v^2+\lambda_3w^2
-\frac{\lambda_1+\lambda_2+\lambda_3}{2\alpha},
\end{equation}
so their geometry is determined by the rank, signature, and trace of \(Q\).

The two relevant degeneracy conditions have distinct meanings. Since
\(\nabla P_2=2Q\mathbf{x}\), a finite singularity occurs precisely when
\begin{equation}
\operatorname{tr}Q=0.
\label{eq:3d_affine}
\end{equation}
For full-rank indefinite \(Q\), this gives a quadratic cone singular at the origin. Conversely,
\begin{equation}
\det Q=0
\label{eq:3d_projective}
\end{equation}
signals a projective degeneracy: the quadratic part loses rank, producing cylindrical or planar limiting surfaces.

Both mechanisms appear along the normalized family
\begin{equation}
\psi_2(t)
=A(t)\bigl(\Phi_{200}+\Phi_{020}\bigr)+t\Phi_{002},
\qquad
A(t)=\sqrt{\frac{1-t^2}{2}},
\qquad -1\leq t\leq1.
\end{equation}
Here
\begin{equation}
Q(t)=\sqrt{2}\alpha\,
\operatorname{diag}\!\bigl(A(t),A(t),t\bigr).
\end{equation}
For \(0<t<1\), the nodal surface is an ellipsoid. At \(t=0\), the rank drops and it becomes the elliptic cylinder
\begin{equation}
\alpha(x^2+y^2)=1.
\end{equation}
For \(-\sqrt{2/3}<t<0\), it is a one-sheet hyperboloid. At
\begin{equation}
t_c=-\sqrt{\frac{2}{3}},
\end{equation}
the trace vanishes and the nodal surface becomes the singular cone
\begin{equation}
x^2+y^2-2z^2=0.
\end{equation}
For \(-1<t<t_c\), the surface is a two-sheet hyperboloid, while \(t=\pm1\) gives two parallel planes,
\begin{equation}
z=\pm\frac{1}{\sqrt{2\alpha}}.
\end{equation}
The \(N=2\) shell provides the first three-dimensional example in which the nodal geometry changes at fixed energy: rank loss controls the asymptotic restructuring, whereas the trace-zero condition produces the finite topology-changing singularity.

\bibliographystyle{unsrt}
\bibliography{main}

@book{Hirsch1976,
  author    = {Hirsch, Morris W.},
  title     = {Differential Topology},
  series    = {Graduate Texts in Mathematics},
  volume    = {33},
  publisher = {Springer},
  address   = {New York},
  year      = {1976},
  doi       = {10.1007/978-1-4684-9449-5}
}

@book{Dundas2018,
  author    = {Dundas, Bj{\o}rn Ian},
  title     = {A Short Course in Differential Topology},
  series    = {Cambridge Mathematical Textbooks},
  publisher = {Cambridge University Press},
  year      = {2018},
  doi       = {10.1017/9781108349130}
}

@article{FluehmannHome2020,
  author  = {Fl{\"u}hmann, Christa and Home, Jonathan P.},
  title   = {Direct Characteristic-Function Tomography of Quantum States of the Trapped-Ion Motional Oscillator},
  journal = {Phys. Rev. Lett.},
  volume  = {125},
  pages   = {043602},
  year    = {2020},
  doi     = {10.1103/PhysRevLett.125.043602}
}

@book{FultonAlgebraicCurves,
  author    = {Fulton, William},
  title     = {Algebraic {C}urves: {A}n {I}ntroduction to {A}lgebraic {G}eometry},
  series    = {Mathematics Lecture Note Series},
  volume    = {30},
  publisher = {W. A. Benjamin},
  address   = {New York},
  year      = {1969},
  isbn      = {978-0-8053-3080-9}
}

@book{FischerPlaneCurves,
  author     = {Fischer, Gerd},
  title      = {Plane {A}lgebraic {C}urves},
  translator = {Kay, Leslie},
  series     = {Student Mathematical Library},
  volume     = {15},
  publisher  = {American Mathematical Society},
  address    = {Providence, RI},
  year       = {2001},
  isbn       = {978-0-8218-2122-0},
  doi        = {10.1090/stml/015}
}

@book{MilnorSingularPoints,
  author    = {Milnor, John},
  title     = {Singular {P}oints of {C}omplex {H}ypersurfaces},
  series    = {Annals of Mathematics Studies},
  volume    = {61},
  publisher = {Princeton University Press},
  address   = {Princeton, NJ},
  year      = {1968},
  isbn      = {978-0-691-08065-9},
  doi       = {10.1515/9781400881819}
}

@article{SchuergerEngel2023Nodes,
  author  = {Sch{\"u}rger, Peter and Engel, Volker},
  title   = {On the relation between nodal structures in quantum wave functions and particle correlation},
  journal = {AIP Advances},
  volume  = {13},
  number  = {12},
  pages   = {125307},
  year    = {2023},
  month   = dec,
  doi     = {10.1063/5.0180004}
}

@article{SchuergerEngel2023Entropy,
  author  = {Sch{\"u}rger, Peter and Engel, Volker},
  title   = {Differential {S}hannon Entropies Characterizing Electron--Nuclear Dynamics and Correlation: Momentum-Space Versus Coordinate-Space Wave Packet Motion},
  journal = {Entropy},
  volume  = {25},
  number  = {7},
  pages   = {970},
  year    = {2023},
  month   = jun,
  doi     = {10.3390/e25070970}
}

@book{CourHil,
  author    = {Courant, Richard and Hilbert, David},
  title     = {Methods of Mathematical Physics},
  volume    = {1},
  edition   = {First English},
  publisher = {Interscience Publishers},
  address   = {New York},
  year      = {1966}
}

@article{CharronL2025,
  author  = {Charron, Philippe and L{\'e}na, Corentin},
  title   = {Pleijel's {T}heorem for {S}chr{\"o}dinger {O}perators},
  journal = {Annales Henri Poincar{\'e}},
  volume  = {26},
  number  = {3},
  pages   = {759--786},
  year    = {2025},
  month   = mar,
  doi     = {10.1007/s00023-024-01536-w}
}

@article{VladimirIArnold_1989,
  author  = {Arnol'd, Vladimir I. and Vishik, M. I. and Il'yashenko, Yu. S. and Kalashnikov, A. S. and Kondrat'ev, V. A. and Kruzhkov, S. N. and Landis, E. M. and Millionshchikov, V. M. and Oleinik, O. A. and Filippov, A. F. and Shubin, M. A.},
  title   = {Some unsolved problems in the theory of differential equations and mathematical physics},
  journal = {Russian Mathematical Surveys},
  volume  = {44},
  number  = {4},
  pages   = {157--171},
  year    = {1989},
  month   = aug,
  doi     = {10.1070/RM1989v044n04ABEH002139}
}

@article{VladimirIArnold_2011,
  author  = {Arnold, Vladimir I.},
  title   = {Topological properties of eigenoscillations in mathematical physics},
  journal = {Proceedings of the Steklov Institute of Mathematics},
  volume  = {273},
  pages   = {25--34},
  year    = {2011},
  month   = jul,
  doi     = {10.1134/S0081543811040031}
}

@article{Karp_1989,
  author  = {Karpushkin, V. N.},
  title   = {Topology of the zeros of eigenfunctions},
  journal = {Functional Analysis and Its Applications},
  volume  = {23},
  number  = {3},
  pages   = {218--220},
  year    = {1989},
  month   = jul,
  doi     = {10.1007/BF01079529}
}

@article{PhysRevA.50.3065,
  author  = {Y{\'a}{\~n}ez, R. J. and Van Assche, W. and Dehesa, J. S.},
  title   = {Position and momentum information entropies of the {$D$}-dimensional harmonic oscillator and hydrogen atom},
  journal = {Physical Review A},
  volume  = {50},
  number  = {4},
  pages   = {3065--3079},
  year    = {1994},
  month   = oct,
  doi     = {10.1103/PhysRevA.50.3065}
}

@article{DJTI2020,
  author  = {Dehesa, J. S. and Toranzo, I. V.},
  title   = {Dispersion and entropy-like measures of multidimensional harmonic systems: application to {R}ydberg states and high-dimensional oscillators},
  journal = {The European Physical Journal Plus},
  volume  = {135},
  number  = {9},
  pages   = {721},
  year    = {2020},
  month   = sep,
  doi     = {10.1140/epjp/s13360-020-00736-7}
}

@misc{BH2015,
  author        = {B{\'e}rard, Pierre and Helffer, Bernard},
  title         = {Some nodal properties of the quantum harmonic oscillator in two dimensions},
  year          = {2015},
  eprint        = {1506.02374},
  archiveprefix = {arXiv},
  primaryclass  = {math.AP},
  note          = {arXiv:1506.02374v2},
  doi           = {10.48550/arXiv.1506.02374},
  url           = {https://arxiv.org/abs/1506.02374}
}

@article{10.1093/imrn/rny290,
  author  = {Beck, Thomas and Hanin, Boris},
  title   = {Level Spacings and Nodal Sets at Infinity for Radial Perturbations of the Harmonic Oscillator},
  journal = {International Mathematics Research Notices},
  volume  = {2021},
  number  = {7},
  pages   = {5007--5036},
  year    = {2021},
  month   = apr,
  doi     = {10.1093/imrn/rny290}
}

@article{VMajerník_1996,
  author  = {Majern{\'i}k, V. and Opatrn{\'y}, T.},
  title   = {Entropic uncertainty relations for a quantum oscillator},
  journal = {Journal of Physics A: Mathematical and General},
  volume  = {29},
  number  = {9},
  pages   = {2187--2197},
  year    = {1996},
  month   = may,
  doi     = {10.1088/0305-4470/29/9/029}
}

@article{Wehner_2010,
  author  = {Wehner, Stephanie and Winter, Andreas},
  title   = {Entropic uncertainty relations---a survey},
  journal = {New Journal of Physics},
  volume  = {12},
  number  = {2},
  pages   = {025009},
  year    = {2010},
  month   = feb,
  doi     = {10.1088/1367-2630/12/2/025009}
}

@article{Forbes2021,
  author  = {Forbes, Andrew and de Oliveira, Michael and Dennis, Mark R.},
  title   = {Structured light},
  journal = {Nature Photonics},
  volume  = {15},
  pages   = {253--262},
  year    = {2021},
  doi     = {10.1038/s41566-021-00780-4}
}

@article{Butow2024,
  author  = {B{\"u}tow, Johannes and Eismann, J{\"o}rg S. and Sharma, Varun and Brandm{\"u}ller, Dorian and Banzer, Peter},
  title   = {Generating free-space structured light with programmable integrated photonics},
  journal = {Nature Photonics},
  volume  = {18},
  pages   = {243--249},
  year    = {2024},
  doi     = {10.1038/s41566-023-01354-2}
}

@article{Jeon2024,
  author  = {Jeon, Honggi and Kang, Jiyong and Kim, Jaeun and Choi, Wonhyeong and Kim, Kyunghye and Kim, Taehyun},
  title   = {Experimental realization of entangled coherent states in two-dimensional harmonic oscillators of a trapped ion},
  journal = {Scientific Reports},
  volume  = {14},
  number  = {1},
  pages   = {6847},
  year    = {2024},
  month   = mar,
  doi     = {10.1038/s41598-024-57391-6}
}

\end{document}


\title{Supplemental material for: Nodal algebraic curves and entropy diagnostics in degenerate two-dimensional harmonic-oscillator shells}

\author{C. A. Escobar, H. Olivares-Pilon, and A. M. Escobar-Ruiz\\[0.5em]
\small Departamento de F\'{i}sica, Universidad Aut\'onoma Metropolitana Unidad Iztapalapa,\\
\small San Rafael Atlixco 186, 09340 Ciudad de M\'exico, M\'exico\\[0.5em]
\small Corresponding author: \href{mailto:carlos.escobar@xanum.uam.mx}{carlos.escobar@xanum.uam.mx}}
\date{}

\maketitle

\noindent\textbf{Keywords:} nodal curves, harmonic oscillator, entropy diagnostics, degenerate eigenspaces, Hermite--Gaussian modes

\vspace{0.4cm}
\noindent This Supplemental Material derives the separable-endpoint
structure of the representative fixed-shell family for \(N\ge2\), including the
nodal-domain count, the factorization of $S_{\rm dom}$, and the vanishing
of $I(x;y)$. It then gives the explicit Hermite-constrained nodal curves
for the oscillator shells $N=1,2,3,4$, obtains the projective discriminant
and finite-affine resultant of the constrained cubic shell, and details
the numerical implementation of the $N=3$ nodal geometries and entropy
diagnostics, including the critical-value test for finite singularities.
Finally, it analyzes the two lowest degenerate doublets of the
two-dimensional infinite square well, illustrating both interior branch
reconnections and boundary-driven changes of nodal topology, together
with their nodal-domain entropies.

\section{Endpoint structure of the representative family}
\label{app:generalN-endpoint}

For the representative family with \(N\ge2\),
\begin{equation}
\begin{aligned}
\psi_N^{(t)}(x,y)
&=
\sqrt{1-t^2}\,\frac{\Phi_{N,0}(x,y)+\Phi_{0,N}(x,y)}{\sqrt2}
+t\,\Phi_{n_+,n_-}(x,y),\\
n_-&=\Big\lfloor\frac N2\Big\rfloor,
\qquad
n_+=\Big\lceil\frac N2\Big\rceil.
\end{aligned}
\label{eq:app-general-family}
\end{equation}
the endpoint $t=1$ is
\begin{equation}
\psi_N^{(1)}(x,y)=\Phi_{n_+,n_-}(x,y)=\phi_{n_+}(x)\phi_{n_-}(y).
\label{eq:app-endpoint-state}
\end{equation}
Since
\begin{equation}
\phi_n(x)=C_n\,H_n(\sqrt{\alpha}\,x)e^{-\alpha x^2/2},
\qquad
C_n\neq 0,
\label{eq:app-hermite-form}
\end{equation}
the Gaussian factor never vanishes, and the nodal set is determined entirely by
the zeros of $H_{n_+}(\sqrt{\alpha}\,x)$ and $H_{n_-}(\sqrt{\alpha}\,y)$. If
$\xi_1,\dots,\xi_{n_+}$ are the zeros of $H_{n_+}$ and
$\eta_1,\dots,\eta_{n_-}$ are the zeros of $H_{n_-}$, then
\begin{equation}
\mathcal N(\psi_N^{(1)})
=
\bigcup_{j=1}^{n_+}\Bigl\{x=\frac{\xi_j}{\sqrt{\alpha}}\Bigr\}\times\mathbb R
\;\cup\;
\mathbb R\times\bigcup_{k=1}^{n_-}\Bigl\{y=\frac{\eta_k}{\sqrt{\alpha}}\Bigr\}.
\label{eq:app-endpoint-nodal-set}
\end{equation}
Hence the plane is partitioned into
\begin{equation}
\nu_N(1)=(n_++1)(n_-+1)
=
\begin{cases}
(\ell+1)^2, & N=2\ell,\\[1mm]
(\ell+1)(\ell+2), & N=2\ell+1,
\end{cases}
\label{eq:app-endpoint-count}
\end{equation}
nodal domains.

Let
\[
I_j^{(n_+)}=\Bigl(\frac{\xi_{j-1}}{\sqrt{\alpha}},\frac{\xi_j}{\sqrt{\alpha}}\Bigr),
\qquad
J_k^{(n_-)}=\Bigl(\frac{\eta_{k-1}}{\sqrt{\alpha}},\frac{\eta_k}{\sqrt{\alpha}}\Bigr),
\]
with $\xi_0=\eta_0=-\infty$ and
$\xi_{n_++1}=\eta_{n_-+1}=+\infty$. Then each nodal domain is a rectangle
\[
\Omega_{jk}=I_j^{(n_+)}\times J_k^{(n_-)}.
\]
Because the density factorizes,
\begin{equation}
\rho_N^{(1)}(x,y)=|\phi_{n_+}(x)|^2\,|\phi_{n_-}(y)|^2,
\label{eq:app-endpoint-density}
\end{equation}
its weight on $\Omega_{jk}$ is
\begin{equation}
\mathbb P(\Omega_{jk})
=
\iint_{\Omega_{jk}}\rho_N^{(1)}(x,y)\,dx\,dy
=
p_j^{(n_+)}q_k^{(n_-)},
\label{eq:app-endpoint-prob}
\end{equation}
where
\begin{equation}
p_j^{(n_+)}=\int_{I_j^{(n_+)}}|\phi_{n_+}(x)|^2\,dx,
\qquad
q_k^{(n_-)}=\int_{J_k^{(n_-)}}|\phi_{n_-}(y)|^2\,dy.
\label{eq:app-endpoint-weights}
\end{equation}
Therefore the nodal-domain entropy factorizes:
\begin{align}
S_{\mathrm{dom}}(1)
&=
-\sum_{j,k}p_j^{(n_+)}q_k^{(n_-)}
\ln\!\bigl(p_j^{(n_+)}q_k^{(n_-)}\bigr)
\nonumber\\
&=
-\sum_j p_j^{(n_+)}\ln p_j^{(n_+)}
-\sum_k q_k^{(n_-)}\ln q_k^{(n_-)}
\nonumber\\
&=
S_{\mathrm{dom}}^{1\mathrm D}(n_+)+S_{\mathrm{dom}}^{1\mathrm D}(n_-),
\label{eq:app-endpoint-Sdom}
\end{align}
with
\begin{equation}
S_{\mathrm{dom}}^{1\mathrm D}(n)
=
-\sum_{\ell=1}^{n+1}p_\ell^{(n)}\ln p_\ell^{(n)}.
\label{eq:app-1D-Sdom}
\end{equation}
Finally, since the density is a product measure, the Cartesian mutual
information vanishes:
\begin{equation}
I(x;y)\big|_{t=1}=0.
\label{eq:app-endpoint-MI}
\end{equation}

\section{Low-shell nodal curves for
\texorpdfstring{\(N=1,2,3,4\)}{N=1,2,3,4}}
\label{appC}

This section records the algebraic curve classes arising in the first few
fixed oscillator shells. The aim is not to classify all real curves of a given
degree, but to identify the restricted curve families generated by
fixed-shell oscillator superpositions.

We use dimensionless variables
\[
\xi=\sqrt{\alpha}\,x,\qquad \eta=\sqrt{\alpha}\,y,
\]
and write a general real state in shell \(N\) as
\[
\psi_N(x,y)=\sum_{n=0}^{N}c_n\Phi_{n,N-n}(x,y),
\qquad
\sum_{n=0}^{N}c_n^2=1 .
\]
Since
\[
\phi_n(x)\propto H_n(\xi)e^{-\xi^2/2},
\]
the nodal set is determined, up to an irrelevant nonzero factor, by the
dimensionless polynomial
\[
Q_N(\xi,\eta)
=
\sum_{n=0}^{N}
c_n
\sqrt{\binom{N}{n}}\,
H_n(\xi)H_{N-n}(\eta).
\]
Thus the physical nodal set is equivalent, under the rescaling
\((x,y)\mapsto(\xi,\eta)\), to
\[
\widehat{\mathcal N}_N
=
\{(\xi,\eta)\in\mathbb R^2:Q_N(\xi,\eta)=0\}.
\]

The parity of the Hermite polynomials gives
\[
Q_N(-\xi,-\eta)=(-1)^NQ_N(\xi,\eta).
\]
Hence odd shells produce odd algebraic curves, while even shells produce even
ones. This parity restriction is one reason fixed-shell nodal curves form
proper subfamilies of all real algebraic curves of degree \(N\).

Finite singular points satisfy
\[
Q_N(\xi_c,\eta_c)=0,\qquad
\partial_\xi Q_N(\xi_c,\eta_c)=0,\qquad
\partial_\eta Q_N(\xi_c,\eta_c)=0.
\]
Asymptotic degeneracies are controlled instead by the highest-degree
homogeneous part. If
\[
Q_{N,\mathrm{top}}(r\cos\theta,r\sin\theta)=r^N f_N(\theta),
\]
then repeated asymptotic directions occur when
\[
f_N(\theta)=0,\qquad f_N'(\theta)=0.
\]
This separates finite affine singularities from degeneracies associated with
the line at infinity.

\subsection*{\(N=1\): linear nodal curves}

For \(N=1\),
\[
Q_1(\xi,\eta)=2(c_1\xi+c_0\eta).
\]
Thus the nodal curve is the line
\[
c_1\xi+c_0\eta=0.
\]
Since
\[
\nabla Q_1=(2c_1,2c_0)
\]
is nonzero for every normalized state, the conditions
\[
Q_1=0,\qquad \nabla Q_1=0
\]
cannot be satisfied. The \(N=1\) shell therefore has no nodal bifurcation:
coefficient variation only rotates the nodal line.

\subsection*{\(N=2\): centered conics}

For \(N=2\), an equivalent polynomial is
\[
Q_2(\xi,\eta)
=
\sqrt{2}\,c_2\xi^2
+
2c_1\xi\eta
+
\sqrt{2}\,c_0\eta^2
-
\frac{c_0+c_2}{\sqrt{2}} .
\]
With \(c_2=a\), \(c_1=b\), and \(c_0=c\), this is the dimensionless form of
the quadratic polynomial used in the main text. It can be written as
\[
Q_2(\xi,\eta)
=
\begin{pmatrix}\xi&\eta\end{pmatrix}
M
\begin{pmatrix}\xi\\ \eta\end{pmatrix}
+D,
\]
where
\[
M=
\begin{pmatrix}
\sqrt{2}\,c_2 & c_1\\
c_1 & \sqrt{2}\,c_0
\end{pmatrix},
\qquad
D=-\frac{c_0+c_2}{\sqrt{2}},
\]
and
\[
\det M=2c_0c_2-c_1^2.
\]

The finite affine singular condition is
\[
D=0,
\qquad\text{equivalently}\qquad
c_0+c_2=0.
\]
In this case the conic is homogeneous and generically factors into two
intersecting lines through the origin.

A separate degeneracy occurs when the quadratic part loses rank:
\[
\det M=0,
\qquad\text{equivalently}\qquad
c_1^2=2c_0c_2.
\]
When \(D\neq0\), this gives a projectively degenerate conic, realized in the
affine plane as two parallel lines. Along the symmetric path
\[
c_2=c_0=\sqrt{\frac{1-t^2}{2}},\qquad c_1=t,
\]
this occurs at
\[
t=\frac{1}{\sqrt{2}},
\]
where the closed conic passes through a parallel-line configuration before
becoming hyperbola-type.

Equivalently, the projective conic discriminant factors as
\[
D\,\det M=0,
\]
or
\[
(c_0+c_2)(2c_0c_2-c_1^2)=0.
\]
The first factor gives finite affine singularities; the second gives the
rank-degenerate quadratic part controlling the ellipse-to-hyperbola
transition.

\subsection*{\(N=3\): constrained cubic curves}

For \(N=3\), after division by an irrelevant nonzero factor,
\[
\begin{aligned}
Q_3(\xi,\eta)
={}&
2c_3\xi^3
+
2\sqrt{3}\,c_2\xi^2\eta
+
2\sqrt{3}\,c_1\xi\eta^2
+
2c_0\eta^3
\\
&-
(3c_3+\sqrt{3}\,c_1)\xi
-
(\sqrt{3}\,c_2+3c_0)\eta .
\end{aligned}
\]
Thus the \(N=3\) shell produces an odd cubic of the restricted form
\[
Q_3(\xi,\eta)
=
A\xi^3+B\xi^2\eta+C\xi\eta^2+D\eta^3+E\xi+F\eta,
\]
where the linear and cubic coefficients are not independent. Hence the shell
does not realize an arbitrary real cubic.

Since \(Q_3\) is odd,
\[
Q_3(0,0)=0,
\]
so the origin always lies on the nodal curve. It is regular unless
\[
3c_3+\sqrt{3}\,c_1=0,
\qquad
\sqrt{3}\,c_2+3c_0=0.
\]
More generally, finite singularities are obtained from
\[
Q_3=0,\qquad
\partial_\xi Q_3=0,\qquad
\partial_\eta Q_3=0.
\]
Eliminating \((\xi,\eta)\) gives the finite cubic discriminant surface in
coefficient space.

The cubic shell also has asymptotic degeneracies controlled by
\[
Q_{3,\mathrm{top}}(\xi,\eta)
=
2c_3\xi^3
+
2\sqrt{3}\,c_2\xi^2\eta
+
2\sqrt{3}\,c_1\xi\eta^2
+
2c_0\eta^3.
\]
Repeated real zeros of
\[
Q_{3,\mathrm{top}}(\cos\theta,\sin\theta)=0
\]
change the arrangement of unbounded branches and should be distinguished from
finite singularities.

Product states give useful reducible checkpoints. For example,
\[
\Phi_{21}:\qquad
Q_3(\xi,\eta)\propto H_2(\xi)H_1(\eta)
\propto \eta(2\xi^2-1),
\]
so the nodal set consists of
\[
\eta=0,\qquad \xi=\pm\frac{1}{\sqrt{2}}.
\]

\subsection*{\(N=4\): constrained quartic curves}

For \(N=4\), the shell gives an even quartic. A compact expression is
\[
\begin{aligned}
Q_4(\xi,\eta)
={}&
c_4H_4(\xi)
+
2c_3H_3(\xi)H_1(\eta)
+
\sqrt{6}\,c_2H_2(\xi)H_2(\eta)
\\
&+
2c_1H_1(\xi)H_3(\eta)
+
c_0H_4(\eta).
\end{aligned}
\]
Using
\[
\begin{aligned}
H_1(z)&=2z,\qquad
H_2(z)=4z^2-2,\\
H_3(z)&=8z^3-12z,\qquad
H_4(z)=16z^4-48z^2+12,
\end{aligned}
\]
and dividing by an irrelevant nonzero factor gives
\[
\begin{aligned}
Q_4(\xi,\eta)
={}&
4c_4\xi^4
+
8c_3\xi^3\eta
+
4\sqrt{6}\,c_2\xi^2\eta^2
\\
&+
8c_1\xi\eta^3
+
4c_0\eta^4
-
(12c_4+2\sqrt{6}\,c_2)\xi^2
\\
&-
12(c_3+c_1)\xi\eta
-
(2\sqrt{6}\,c_2+12c_0)\eta^2
\\
&+
3(c_4+c_0)
+
\sqrt{6}\,c_2 .
\end{aligned}
\]
Thus \(N=4\) produces a five-parameter family of even quartics:
\[
Q_4(-\xi,-\eta)=Q_4(\xi,\eta).
\]
These quartics contain fourth-degree, second-degree, and constant terms, but
no cubic or linear terms. Therefore the \(N=4\) shell is again a restricted
algebraic family.

Possible nodal geometries include smooth quartics with closed oval
components, unbounded quartic branches, reducible conic--conic
configurations, products involving symmetric pairs of lines, and singular
quartics with double points or tangencies. Finite singularities are found from
\[
Q_4=0,\qquad
\partial_\xi Q_4=0,\qquad
\partial_\eta Q_4=0,
\]
while the arrangement of unbounded branches is controlled by
\[
Q_{4,\mathrm{top}}(\xi,\eta)
=
4c_4\xi^4
+
8c_3\xi^3\eta
+
4\sqrt{6}\,c_2\xi^2\eta^2
+
8c_1\xi\eta^3
+
4c_0\eta^4 .
\]

The product state \(\Phi_{22}\) gives
\[
Q_4(\xi,\eta)\propto H_2(\xi)H_2(\eta)
\propto (2\xi^2-1)(2\eta^2-1).
\]
Hence the nodal set is the rectangular grid
\[
\xi=\pm\frac{1}{\sqrt{2}},
\qquad
\eta=\pm\frac{1}{\sqrt{2}},
\]
which partitions the plane into
\[
(2+1)(2+1)=9
\]
nodal domains. This agrees with the separable-endpoint count
\[
\nu_N=(n_++1)(n_-+1)
\]
for \(N=4\), where \(n_+=n_-=2\).

\subsection*{Relevance to the entropy diagnostics}

The low-shell hierarchy is
\[
\begin{aligned}
N=1:&\ \text{lines},\\
N=2:&\ \text{centered conics},\\
N=3:&\ \text{constrained odd cubics},\\
N=4:&\ \text{constrained even quartics}.
\end{aligned}
\]
This hierarchy is the geometric input behind the entropy diagnostics used in
the main text. Away from finite singularities and repeated asymptotic
directions, the nodal set deforms by ambient isotopy and the nodal weights
vary smoothly. Changes in \(S_{\mathrm{dom}}\) associated with
nodal-topology restructuring occur when the nodal partition changes, whereas
the global entropies can remain regular because the nodal set has measure zero
with respect to the Gaussian-weighted probability density.

\subsection{Cubic-shell discriminants for
\texorpdfstring{$N=3$}{N=3}}
\label{app:cubic-discriminants}

The \(N=3\) shell does not generate arbitrary real cubics. Its polynomial has
the Hermite-constrained odd form
\begin{equation}
P_3(x,y)
=
A x^3+B x^2y+Cxy^2+Dy^3+E x+F y ,
\label{eq:app-cubic-P3}
\end{equation}
with the linear coefficients fixed by the cubic part:
\begin{equation}
E=-\frac{3A+C}{2\alpha},
\qquad
F=-\frac{B+3D}{2\alpha}.
\label{eq:app-cubic-linear-constraint}
\end{equation}
Equivalently, if
\begin{equation}
H_3(x,y)=Ax^3+Bx^2y+Cxy^2+Dy^3 ,
\end{equation}
then
\begin{equation}
P_3
=
H_3-\frac{1}{4\alpha}\nabla^2 H_3 ,
\label{eq:P3-H3-laplacian}
\end{equation}
where
\begin{equation}
\nabla^2=\partial_x^2+\partial_y^2
\end{equation}
is the two-dimensional Laplacian. We reserve the symbol
$\Delta_\infty$ for the projective discriminant of the leading
binary cubic.
Thus the cubic-shell nodal set is controlled by the highest homogeneous part
\(H_3\), together with the lower-order Hermite correction.

There are two distinct degeneracy mechanisms. The asymptotic directions are
the real projective roots of \(H_3=0\). They become non-simple when the binary
cubic discriminant vanishes:
\begin{equation}
\Delta_{\infty}
=
B^2C^2
-4AC^3
-4B^3D
-27A^2D^2
+18ABCD
=0 .
\label{eq:app-cubic-projective-discriminant}
\end{equation}
This is the projective, or asymptotic, discriminant. It controls changes in
the arrangement of unbounded branches.

Finite affine singularities instead require
\begin{equation}
P_3(x_c,y_c)=0,
\qquad
\nabla P_3(x_c,y_c)=0 .
\label{eq:app-cubic-finite-singularity}
\end{equation}
Define
\[
p=3A+C,\qquad q=B+3D,\qquad \ell(x,y)=px+qy .
\]
Using Euler's identity for \(H_3\), any finite singular point must satisfy
\[
H_3(x_c,y_c)=0,
\qquad
\ell(x_c,y_c)=0 .
\]
Hence finite singular candidates occur only when \(H_3\) and \(\ell\) have a
common projective root, equivalently when
\begin{equation}
R_{\mathrm{fin}}
=
Aq^3
-Bpq^2
+Cp^2q
-Dp^3
=0 .
\label{eq:app-cubic-finite-resultant}
\end{equation}

This resultant provides a necessary algebraic condition for a
finite-singularity candidate. A real affine singularity occurs only on the
sublocus for which the corresponding common root is real and satisfies the
full affine singularity conditions in
Eq.~\eqref{eq:app-cubic-finite-singularity}.

For the three-state path used in the main text,
\[
a(t)=c(t)=\sqrt{\frac{1-t^2}{2}},
\qquad
b(t)=t,
\qquad
d(t)=0,
\]
one has, up to a common nonzero factor,
\[
A=\frac{2}{\sqrt3}\alpha^{3/2}a(t),
\quad
B=2\alpha^{3/2}t,
\quad
C=2\alpha^{3/2}a(t),
\quad
D=0 .
\]
The projective discriminant reduces to
\begin{equation}
\Delta_{\infty}
=
C^2(B^2-4AC),
\label{eq:app-path-projective-discriminant}
\end{equation}
and its nontrivial zero is
\begin{equation}
t_{\infty}^2
=
4-2\sqrt3,
\qquad
t_{\infty}\simeq0.732 .
\label{eq:app-path-tinfty}
\end{equation}
This value lies in the same range where Fig.~6 of the main text shows the
closest approach of smooth nodal branches.

The finite resultant has endpoint zeros at \(t=0\) and \(t=1\), corresponding
to the line--ellipse configuration and the separable \(\Phi_{21}\) endpoint,
respectively. It also has an interior reducible candidate at
\begin{equation}
t_{\mathrm{red}}^2
=
\frac{\sqrt3}{1+\sqrt3}
=
\frac{3-\sqrt3}{2},
\qquad
t_{\mathrm{red}}\simeq0.796 .
\label{eq:app-path-tred}
\end{equation}
A direct real check shows that this interior reducible point does not produce
real affine singularities. This is consistent with the critical-value
diagnostic \(\Delta_{\mathrm{crit}}\), which remains nonzero on the interior
branch \(0<t<1\).

Thus, along the path studied in the main text, the observed close branch approach is
controlled by the nearby projective discriminant
\(\Delta_{\infty}=0\), not by a real finite singularity. The real finite
singular configurations occur at the degenerate endpoints.

\section{Numerical implementation}
\label{app:numerics}

This section summarizes the numerical procedure used to generate the
representative $N=3$ nodal geometries and nodal-domain entropy shown in
Figs.~6 and 7 of the main text. All calculations were performed in the
dimensionless oscillator variables
\[
X=\sqrt{\alpha}\,x,\qquad
Y=\sqrt{\alpha}\,y,\qquad
\Pi_x=\frac{p_x}{\sqrt{m\hbar\omega}},\qquad
\Pi_y=\frac{p_y}{\sqrt{m\hbar\omega}}.
\]
We therefore work in oscillator units, with $\alpha=1$, and relabel $X$ and
$Y$ as $x$ and $y$ for notational simplicity. All probability densities are
normalized with respect to the corresponding dimensionless measures, so the
arguments of logarithms appearing in differential entropies are
dimensionless. A change of physical units adds only fixed,
coefficient-independent constants to the differential entropies and does not
affect comparisons along the parameter path. The nodal-domain entropy,
being defined from dimensionless probabilities, is unaffected.

Along the path used in the main text,
\[
a(t)=c(t)=\sqrt{\frac{1-t^2}{2}},\qquad b(t)=t,\qquad d(t)=0,
\]
the cubic polynomial was taken, up to an irrelevant overall factor, as
\begin{equation}
\begin{aligned}
P_3(x,y;t)
&=
\frac{2}{\sqrt3}\alpha^{3/2}a(t)x^3
+2\alpha^{3/2}b(t)x^2y
+2\alpha^{3/2}c(t)xy^2  \\
&\quad
-\sqrt{\alpha}\bigl(\sqrt3\,a(t)+c(t)\bigr)x
-\sqrt{\alpha}\,b(t)y .
\end{aligned}
\label{eq:app-num-P3}
\end{equation}
For the computation of domain weights we used the Gaussian-weighted
polynomial density
\[
w_3(x,y;t)=e^{-\alpha(x^2+y^2)}P_3(x,y;t)^2,
\]
and normalized the weights by dividing by the total integral over the computational domain.

The nodal plots in Fig.~6 of the main text show the zero contour of \(P_3\) on
\([-3.2,3.2]^2\) for
\[
t=0.10,\quad 0.70,\quad 0.85,\quad 1.00 .
\]
The background shading represents the value of the polynomial field \(P_3\),
clipped and rescaled for visual contrast; it is not the probability density.

For the entropy curve in Fig.~7 of the main text, the interior values
\(0<t<1\) were computed on a Cartesian grid on \([-L,L]^2\), with \(L=8\) and
\(n=180\) subdivisions in each direction. At each grid point, the sign of
\(P_3\) was recorded using the tolerance \(\varepsilon=10^{-12}\):
\[
\sigma_{ij}
=
\begin{cases}
+1, & P_3(x_i,y_j;t)>\varepsilon,\\
-1, & P_3(x_i,y_j;t)<-\varepsilon,\\
0, & |P_3(x_i,y_j;t)|\leq \varepsilon .
\end{cases}
\]
Grid points with \(\sigma_{ij}=0\) were treated as nodal points. Connected
nodal domains were then identified as nearest-neighbor connected components of
the sign field \(\sigma_{ij}=\pm1\). For each component \(C\), the unnormalized
weight was computed by
\begin{equation}
w_C(t)
=
\sum_{(i,j)\in C}
w_3(x_i,y_j;t)\,\Delta x\,\Delta y .
\label{eq:app-num-weight}
\end{equation}
After discarding components with \(w_C<10^{-14}\), the weights were normalized,
\[
p_C(t)=\frac{w_C(t)}{\sum_{C'}w_{C'}(t)},
\]
and the nodal-domain entropy was evaluated as
\begin{equation}
S_{\mathrm{dom}}(t)
=
-\sum_C p_C(t)\ln p_C(t).
\label{eq:app-num-Sdom}
\end{equation}

The endpoints were treated separately from their analytic nodal
configurations. At \(t=0\),
\begin{equation}
P_3(x,y;0)
\propto
x\left[
\frac{2}{\sqrt3}\alpha x^2
+
2\alpha y^2
-
(\sqrt3+1)
\right],
\label{eq:app-num-t0}
\end{equation}
so the nodal set is the union of the line \(x=0\) and the ellipse
\[
\frac{2}{\sqrt3}\alpha x^2+2\alpha y^2=\sqrt3+1 .
\]
At \(t=1\),
\begin{equation}
P_3(x,y;1)\propto y(2\alpha x^2-1),
\label{eq:app-num-t1}
\end{equation}
so the nodal set consists of \(y=0\) and
\(x=\pm1/\sqrt{2\alpha}\), giving six nodal domains. Since this endpoint is the
separable product state \(\Phi_{21}\), its nodal-domain entropy factorizes:
\begin{equation}
S_{\mathrm{dom}}(1)
=
S_{\mathrm{dom}}^{1D}(2)+S_{\mathrm{dom}}^{1D}(1).
\label{eq:app-num-t1-Sdom}
\end{equation}

As checks, the endpoint factorizations reproduce the plotted nodal
geometries, the separable endpoint satisfies $I(x;y)=0$, and the
Gaussian tail outside $[-8,8]^2$ is negligible in the units used here.
All entropy curves were stable under changes of the computational
window, grid resolution, quadrature parameters, and nodal cutoff: the
values of $S_{\rm dom}$, $I(x;y)$, and $S_r+S_p$ were unchanged at the
graphical resolution for regular parameter values. Near algebraic
transition points, the nodal topology was assigned from the analytic
degeneracy conditions rather than from the finite grid alone.

\subsection{Critical-value diagnostic}
\label{app:critical-value-diagnostic}

To distinguish an exact finite singularity from a close approach of smooth
nodal branches, we also computed a critical-value diagnostic for the \(N=3\)
path. For each sampled value of \(t\), the real critical points of the cubic
were obtained by solving
\[
\partial_x P_3(x,y;t)=0,
\qquad
\partial_y P_3(x,y;t)=0 .
\]
At such a point, an actual finite singularity of the nodal curve occurs only if
the critical value also vanishes, \(P_3(x_c,y_c;t)=0\). We therefore define
\begin{equation}
\Delta_{\mathrm{crit}}(t)
=
\min_{\nabla P_3(x_c,y_c;t)=0}
\frac{|P_3(x_c,y_c;t)|}
{\|P_3(\cdot,\cdot;t)\|_G},
\label{eq:app-Delta-crit}
\end{equation}
where the Gaussian norm is
\begin{equation}
\|P_3(\cdot,\cdot;t)\|_G
=
\left[
\iint_{\mathbb R^2}
e^{-\alpha(x^2+y^2)}P_3(x,y;t)^2\,dx\,dy
\right]^{1/2}.
\label{eq:app-Gaussian-norm-P3}
\end{equation}
Thus \(\Delta_{\mathrm{crit}}(t)=0\) precisely when the finite singularity
condition
\[
P_3(x_c,y_c;t)=0,
\qquad
\nabla P_3(x_c,y_c;t)=0
\]
is satisfied.

Numerically, the minimum in \eqref{eq:app-Delta-crit} was taken over the real
critical points lying inside the computational window. Along the interior
branch \(0<t<1\), this diagnostic remains nonzero, confirming that the
close-branch regime shown in Fig.~6 of the main text is not an exact finite
singularity. It is instead a close approach of smooth nodal branches.
Small values of \(\Delta_{\mathrm{crit}}\) occur near the degenerate endpoints,
which are treated separately from the regular interior branch.

\section{Lowest degenerate doublets of the two-dimensional infinite square well}
\label{app:square-well}

The two-dimensional infinite square well provides a simple nonoscillator
example of nodal restructuring at fixed energy. Consider a particle confined
to the square
\[
\Omega=(0,L)\times(0,L).
\]
The normalized eigenfunctions and their energies are
\begin{equation}
\Phi_{n_xn_y}(x,y)
=
\frac{2}{L}
\sin\left(\frac{n_x\pi x}{L}\right)
\sin\left(\frac{n_y\pi y}{L}\right),
\qquad
E_{n_xn_y}
=
E_0\left(n_x^2+n_y^2\right),
\label{eq:box-eigenstates}
\end{equation}
where
\[
E_0=\frac{\hbar^2\pi^2}{2mL^2}.
\]
For \(n_x\neq n_y\), the states \(\Phi_{n_xn_y}\) and
\(\Phi_{n_yn_x}\) are degenerate. We consider the first two such doublets,
\begin{equation}
\begin{aligned}
\mathcal S_1
&=
\operatorname{span}\{\Phi_{12},\Phi_{21}\},
& E^{\square}_1&=5E_0,\\
\mathcal S_2
&=
\operatorname{span}\{\Phi_{13},\Phi_{31}\},
& E^{\square}_2&=10E_0.
\end{aligned}
\label{eq:box-doublets}
\end{equation}
Here \(N=1,2\) labels the first two permutation-degenerate doublets and
should not be confused with the oscillator shell number. Unlike the
isotropic oscillator, the square-well spectrum is not organized by
\(n_x+n_y\).

A real state in either doublet can be written as
\begin{equation}
\Psi_N(x,y;a,b)
=
a\,\Phi_{1,N+1}(x,y)
+
b\,\Phi_{N+1,1}(x,y),
\qquad
a^2+b^2=1.
\label{eq:box-superposition}
\end{equation}
Introduce the dimensionless coordinates
\[
X=\frac{\pi x}{L},
\qquad
Y=\frac{\pi y}{L},
\qquad
0<X,Y<\pi.
\]
Using
\[
\sin[(N+1)z]
=
\sin z\,U_N(\cos z),
\]
where \(U_N\) is a Chebyshev polynomial of the second kind,
Eq.~\eqref{eq:box-superposition} becomes
\begin{equation}
\Psi_N
=
\frac{2}{L}\sin X\sin Y
\left[
a\,U_N(\cos Y)+b\,U_N(\cos X)
\right].
\label{eq:box-chebyshev}
\end{equation}
The factors \(\sin X\) and \(\sin Y\) vanish only at the walls. Therefore,
inside the well the nodal curve is determined by
\begin{equation}
a\,U_N(\cos Y)+b\,U_N(\cos X)=0.
\label{eq:box-nodal-curve}
\end{equation}
Thus the mixing coefficients \(a\) and \(b\) change the nodal geometry
without changing the energy.

\subsection{First doublet: \(N=1\)}

For \(N=1\), \(U_1(z)=2z\), and the interior nodal curve is
\begin{equation}
a\cos Y+b\cos X=0.
\label{eq:box-N1}
\end{equation}
With \(u=\cos X\) and \(v=\cos Y\), this is the straight line
\[
bu+av=0
\]
inside the square \((-1,1)\times(-1,1)\). It continuously rotates as the
coefficients vary. Moreover,
\[
\partial_X P_1=-b\sin X,
\qquad
\partial_Y P_1=-a\sin Y,
\]
so \(P_1=\nabla P_1=0\) has no solution in the interior for a normalized
state. The nodal curve is therefore always smooth and divides the well into
two connected domains.

The transformation
\[
(X,Y)\longmapsto(\pi-X,\pi-Y)
\]
interchanges these two domains while leaving \(|\Psi_1|^2\) unchanged.
Consequently, both domains carry probability \(1/2\), and
\begin{equation}
S_{\rm dom}^{(N=1)}=\ln 2
\label{eq:box-N1-entropy}
\end{equation}
for every real superposition. This doublet is the square-well analogue of
the \(N=1\) oscillator shell: coefficient variation changes the orientation
of the node but not its topology or domain weights.

\subsection{Second doublet: \(N=2\)}

For \(N=2\), \(U_2(z)=4z^2-1\). The interior nodal curve becomes
\begin{equation}
4b\cos^2X+4a\cos^2Y-(a+b)=0.
\label{eq:box-N2}
\end{equation}
In the variables \(u=\cos X\) and \(v=\cos Y\), this is the centered conic
\begin{equation}
4bu^2+4av^2=a+b.
\label{eq:box-N2-conic}
\end{equation}
Its geometry depends directly on the relative magnitude and sign of the
superposition coefficients.

For \(ab>0\), an overall sign can be chosen so that \(a,b>0\), and
Eq.~\eqref{eq:box-N2-conic} is an ellipse. It lies completely inside the
physical square when
\begin{equation}
\frac{1}{3}<\frac{a}{b}<3.
\label{eq:box-ellipse-range}
\end{equation}
In this regime, the nodal set is a closed curve and separates the well into
two domains. At \(a=b\), for example,
\[
u^2+v^2=\frac{1}{2}.
\]
When \(a=3b\) or \(b=3a\), the ellipse touches the wall. Outside the range
in Eq.~\eqref{eq:box-ellipse-range}, its intersection with the well consists
of two open nodal arcs, which divide the square into three domains. This is
a boundary-driven transition: the nodal topology changes when the curve
touches the hard wall, even though no singular point appears in the
interior.

For \(ab<0\), the conic is hyperbola-type. An interior singularity occurs
when
\begin{equation}
a+b=0.
\label{eq:box-N2-singular}
\end{equation}
At this point,
\[
v^2-u^2=0,
\]
and the nodal set reduces to the two diagonals
\begin{equation}
Y=X,
\qquad
Y=\pi-X.
\label{eq:box-diagonals}
\end{equation}
The diagonals cross at the center of the well and divide it into four
equivalent nodal domains. By symmetry, each domain has probability \(1/4\),
so
\begin{equation}
S_{\rm dom}^{(a=-b)}=\ln 4.
\label{eq:box-diagonal-entropy}
\end{equation}
On either side of this singular configuration, the two branches reconnect
into a smooth hyperbola-type curve.

The separable endpoints provide another exact check. For \(a=1,b=0\), the
state is \(\Phi_{13}\), with the two horizontal nodal lines
\[
Y=\frac{\pi}{3},
\qquad
Y=\frac{2\pi}{3}.
\]
They divide the well into three equal-probability regions. The endpoint
\(a=0,b=1\) gives the corresponding vertical configuration. In both cases,
\begin{equation}
S_{\rm dom}^{\rm sep}=\ln 3.
\label{eq:box-separable-entropy}
\end{equation}

This example shows that the geometric framework is not restricted to
Gaussian--Hermite states. In the square well, the relevant curves are
generated by Chebyshev-constrained trigonometric functions. Finite
singularities still identify branch reconnections, while the hard-wall
boundary introduces an additional mechanism: nodal curves may change their
connectivity by touching the boundary. The nodal-domain entropy distinguishes
the resulting two-, three-, and four-domain configurations at fixed energy.